\documentclass[aps,twocolumn,floatfix,a4]{revtex4}

\usepackage{amsmath}
\usepackage{latexsym}
\usepackage{float}
\usepackage{amssymb}
\usepackage{graphicx}
\usepackage{epsfig}
\usepackage{color}

\definecolor{red}{rgb}{0.0,0.0,0.0}

\definecolor{red}{rgb}{1.0,0.0,0.0}

\begin{document}

\title{Static and dynamic properties of large polymer melts in equilibrium}
\author{Hsiao-Ping Hsu} 
\email{hsu@mpip-mainz.mpg.de}
\affiliation{Max-Planck-Institut f\"ur Polymerforschung, Ackermannweg 10, 55128, Mainz}
\author{Kurt Kremer}
\email{kremer@mpip-mainz.mpg.de}
\affiliation{Max-Planck-Institut f\"ur Polymerforschung, Ackermannweg 10, 55128, Mainz}

\begin{abstract}
We present a detailed study of the static and dynamic behavior of long semiflexible 
polymer chains in a melt. Starting from previously obtained fully equilibrated high molecular 
weight polymer melts [{\it Zhang et al.} ACS Macro Lett. 3, 198 (2014)]
we investigate their static and dynamic scaling behavior as predicted by 
theory. We find that for semiflexible chains in a melt, results of the 
mean square internal distance, the probability distributions of the 
end-to-end distance, and the chain structure factor are well described 
by theoretical predictions for ideal chains.
We examine the motion of monomers and chains by molecular dynamics 
simulations using the ESPResSo++ package. The scaling predictions of the mean 
squared displacement of inner monomers, center of mass, and relations between them based on the Rouse and the reptation theory are verified, and related characteristic relaxation times are determined. Finally we give evidence that the entanglement length $N_{e,PPA}$ as determined by a primitive path analysis (PPA) predicts a plateau modulus, $G_N^0=\frac{4}{5}(\rho k_BT/N_e)$, consistent with stresses obtained from the Green-Kubo 
relation. These comprehensively characterized equilibrium structures, which offer a good compromise between flexibility, small $N_e$, computational efficiency, and small deviations from ideality provide ideal starting states for future non-equilibrium studies.
\end{abstract}

\maketitle

\newpage
\section{Introduction}

  A fundamental property of polymer melts containing long linear
chains is that they are entangled. As the stiffness of chains increases,
the entanglement effect becomes stronger, i.e. the entanglement
length is shorter. Complex topological constraints in polymer 
melts play an essential role for dynamic, and 
rheological properties. For studying such properties and phenomena
in an out-of-equilibrium state it is important to begin with a well characterized 
equilibrium `sample' of very long polymer chains in a melt. It is the purpose
of this study to provide this.

 According to Flory's argument, the excluded volume interactions
become screened~\cite{deGennes1979, Yamakawa1971} when the concentration of polymer
solutions exceed the chain overlap concentration.
Therefore polymer chains in a melt eventually
behave statistically as ideal chains, as if excluded volume effect would no longer be important. 
However, Wittmer and his co-workers~\cite{Wittmer2007,Wittmer2011}
have pointed out that there are noticeable deviations from an
ideal chain behavior due to the incompressibility constraint of the melt.
For fully flexible polymer chains in a melt based on lattice and continuum
models, bond fluctuation model (BFM) and bead-spring model, respectively, such
deviations are indeed seen. This finding is confirmed by a recent
Monte Carlo study of polymer melts using BFM in
Ref.~\cite{Hsu2014} while the deviations are less visible
as the chain stiffness starts to play a role for polymers. 
Therefore, we provide a detailed study of the conformational properties
of long bead-spring polymer chains in a melt as the chain stiffness is 
taken into account, where we especially study to what extent polymer 
chains behave as ideal chains.

It is well known that for short unentangled chains in a melt, the motion
of monomers can be approximately described by the 
Rouse model~\cite{deGennes1979,Doi1986,Kalathi2014,Kalathi2015}.
If the polymer chains become long enough such that the effects of entanglements
start to become important, movements of chains at the intermediate time and length scales
are confined to a tube-like region, created by surrounding
chains and depending on the corresponding
entanglement length $N_e$. The dynamic behavior within this
time frame is well described by the tube model of de Gennes, 
Doi and Edwards~\cite{deGennes1979,Doi1986,Rubinstein2003}.
Each polymer chain is assumed to move back and forth 
(reptation) along the contour of an imaginary tube around the so called primitive path.
Although ample evidence of reptation scaling predictions is given by previous Monte Carlo and
molecular dynamics simulations~\cite{Baumgaertner1981,Kremer1990,Paul1991,Wittmer1992,Kremer1992,Kopf1997},
a complete picture still is lacking. This is mostly due to
the limitations of available equilibrated systems of huge chain length and the
long relaxation times covering several orders of magnitude.

Recently, the authors of Ref.~\cite{Zhang2014}
developed a novel and very efficient methodology for equilibrating high molecular
weight polymer melts through a sequential backmapping
of a soft-sphere coarse-grained model~\cite{Vettorel2010, Zhang2013}
from low resolution to high resolution, and finally the application of molecular dynamics (MD) simulations of
the underlying bead-spring model (see Appendix).
Therefore, a further investigation of the static and dynamic scaling behavior predicted by
theories~\cite{deGennes1979,Yamakawa1971,Doi1986} for huge systems in the highly entangled 
regime has become easily accessible.
Therefore, the aim of this paper is to give a deeper understanding of
static and dynamic behavior of large semiflexible polymer chains in a melt,
and compare our numerical results whenever it is possible to theoretical predictions     
in the literature. 
We mainly focus on polymer melt system containing $n_c=1000$ semiflexible polymer
chains of sizes $N=500$, $1000$, $2000$ 
with the Flory characteristic ratio $C_\infty \sim 2.88$. The chains are modelled as standard
bead-spring chains with a bond bending interactions parameter $k_\theta=1.5$. For details of
the model we refer to the appendix. All results quoted refer to chains with a bending constant of $k_\theta = 1.5$ unless otherwise noted.
Having such big polymer melt systems at hand we have the possibility to analyze the 
linear viscoelasticity as characterized by the stress relaxation modulus,
and estimate the entanglement length $N_e$ from 
the standard expression of the plateau modulus $G_N^0=(4/5)(\rho k_B T/N_e)$.
It is also interesting to check whether $N_e$ is equivalent to 
the estimate of $N_{e,PPA}$ through the primitive
path analysis (PPA)~\cite{Everaers2004}.

    The outline of the paper is as follows: Sec.~II describes the static conformational 
structures of polymer chains in a melt, and compares them to those for ideal chains.
Sec. III describes the motions of polymer chains in a melt at different 
characteristic time scales, and verifies the scaling laws predicted by the
Rouse model, and the reptation theory~\cite{deGennes1979,Doi1986,Rubinstein2003}.
The detailed structure investigation of the primitive path of chains is given
in Sec.~IV. Studies of linear viscoelasticity of polymer melts are given in Sec. V.
Finally, our conclusions are summarized in Sec. VI.

\section{Static properties of equilibrated polymer melts}

\begin{figure}[t]
\begin{center}
\includegraphics[width=0.30\textwidth,angle=270]{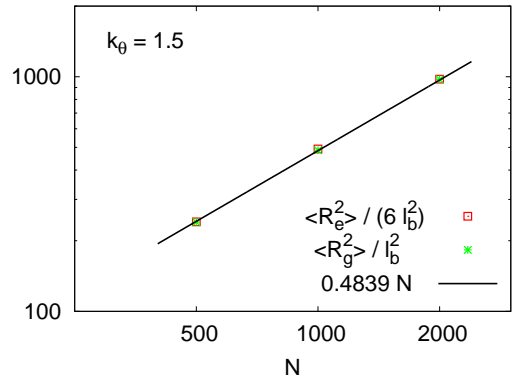} 
\caption{Log-log plot of rescaled mean square end-to-end distance 
$\langle R_e^2 \rangle/(6 \ell_b^2)$ and
gyration radius $\langle R_g^2 \rangle/\ell_b^2$ versus
$N$. The straight line indicates a fit of the power law,
$\langle R_e^2 \rangle \propto \langle R_g^2 \rangle \propto N^{2\nu}$
with $\nu=1/2$ for ideal chains, to the data.}
\label{fig-ReRg2}
\end{center}
\end{figure}

  Let us first look at the estimates of the mean square end-to-end distance
and the mean square radius of gyration given by
\begin{equation}
    \langle R_e^2 \rangle = \frac{1}{n_c} \sum_{i=1}^{n_c}
 \langle (\vec{r}_{i,N} - \vec{r}_{i,1})^2 \rangle \,,
\label{eq-Re2}
\end{equation}
and 
\begin{eqnarray}
 \langle  R_g^2 \rangle = \frac{1}{n_cN} \sum_{i=1}^{n_c} \left \langle 
\sum_{j=1}^N (\vec{r}_{i,j}-\vec{r}_{i,{\rm c.m.}})^2 \right \rangle  
\,, \\
 \textrm{with} \quad \vec{r}_{i,{\rm c.m.}}=\frac{1}{N}\sum_{j=1}^N \vec{r}_{i,j} \nonumber
\label{eq-Rg2}
\end{eqnarray}
where $\vec{r}_{i,j}$ is the position of monomer $j$ of chain number $i$ while  $\vec{r}_{i,{\rm c.m.}}$ is the 
center of mass (c.m.) of the $i$-th polymer chain in a melt, and 
the average $\langle \ldots \rangle$ includes an averaging over all
independent equilibrated configurations.
Results of $\langle R_e^2\rangle /(6\ell_b^2)$ and 
$\langle R_g^2 \rangle/\ell_b^2$ plotted versus $N$
are shown in Fig.~\ref{fig-ReRg2} 
for polymer melts
containing $n_c=1000$ chains of sizes $N=500$, $1000$, and $2000$. Here the root-mean square bond length
$\ell_b=\mid \vec{b}^2 \mid^{1/2} \approx 0.964$.
We see that $\langle R_e^2 \rangle \propto \langle R_g^2 \rangle
\propto N$, and $\langle R_e^2 \rangle /\langle R_g^2 \rangle \approx 6$
as one would expect for ideal chains.

 The conformational behavior of individual polymer chains of size $N$ in
a melt can also be described by
the probability distributions of end-to-end distance $\vec{R}_e$
and gyration radius $R_g$,
$P_N(\vec{R}_e)$ and $P_N(R_g)$, respectively.
For ideal chains where $\langle R_e^2 \rangle \propto N\ell_b^2$, 
the probability distribution of $\vec{R}_e$ is
a Gaussian distribution,
\begin{equation}
    P_N(\vec{R}_e)=\left(\frac{3}{2\pi \langle R_e^2 \rangle}\right)^{3/2}
\exp \left(-\frac{3\vec{R}_e^2}{2 \langle R_e^2 \rangle} \right) \,.
\label{eq-pRe}
\end{equation}
Although there exists an exact theoretical 
prediction~\cite{Yamakawa1971,Fujita1970,Denton2002} for the probability distribution of $R_g$ it is much more complicated
to evaluate.
However, it has been checked~\cite{Vettorel2010,Hsu2014,Hsu2014m}
that the formula suggested by Lhuillier~\cite{Lhuillier1988} for
polymer chains under good solvent conditions in $d$-dimensions is still
a good approximation for ideal chains ($\nu = 1/2$), i.e.,
\begin{equation}
 P_N(R_g) \sim \exp \left[-a_1 \left(\frac{\ell_b N^\nu}{R_g}\right)^{\alpha d}
-a_2 \left(\frac{R_g}{\ell_b N^\nu} \right)^\delta \right]
\label{eq-pRg}
\end{equation}
where $a_1$ and $a_2$ are (non-universal) constants, and the exponents $\alpha$ and
$\delta$ are linked to the space dimension $d$ and the Flory exponent $\nu$ by
$\alpha=(\nu d-1)^{-1}$ and $\delta=(1-\nu)^{-1}$. Here $(1+\alpha)$
is the des Cloizeaux exponent~\cite{Cloizeaux1975} for the osmotic pressure of
a semidilute polymer solution, and $\delta$ is the Fisher exponent~\cite{Fisher1966}
characterizing the end-to-end distance distribution.

\begin{figure}[t!]
\begin{center}
(a)\includegraphics[width=0.30\textwidth,angle=270]{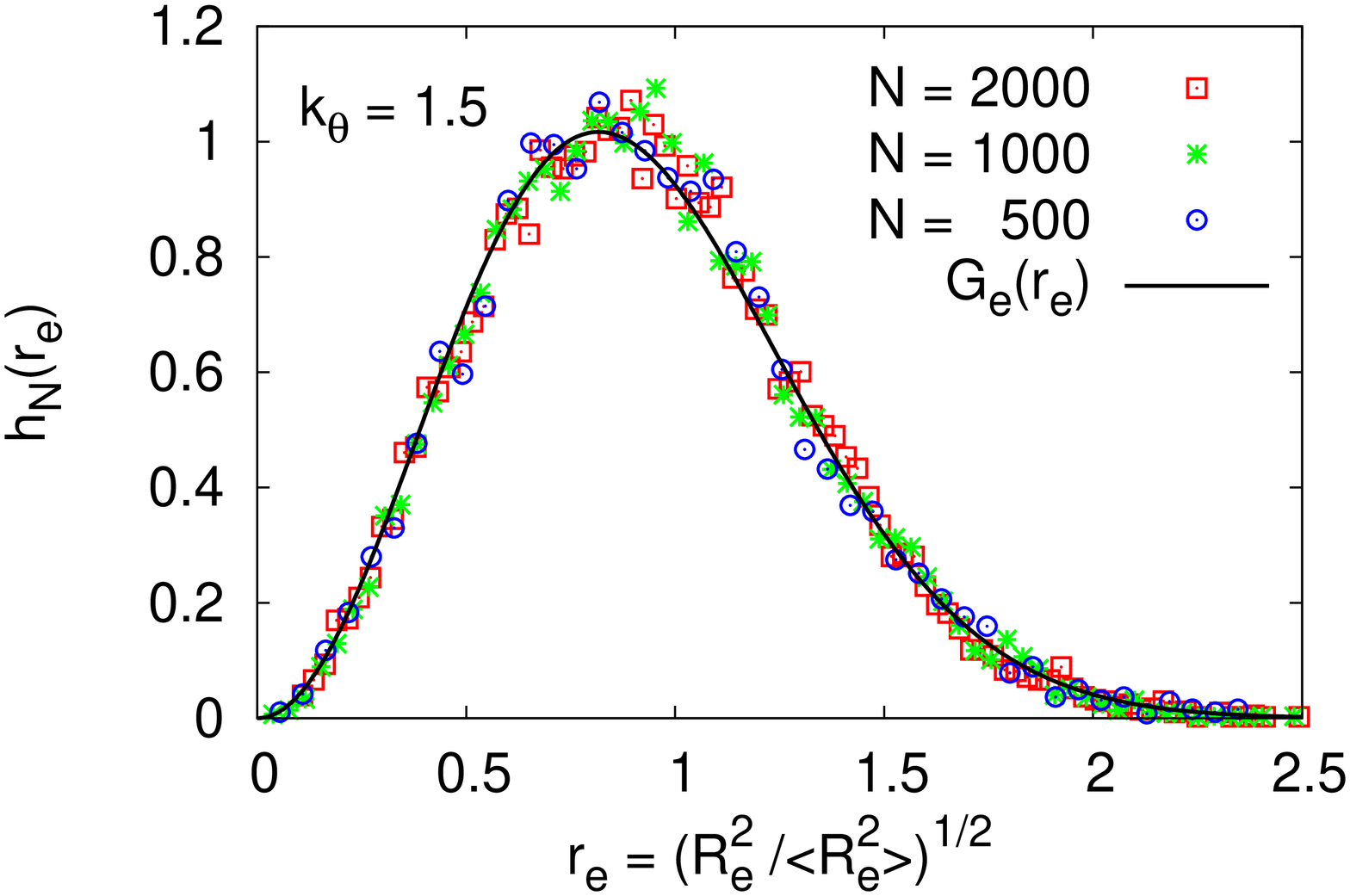} \hspace{0.4cm}
(b)\includegraphics[width=0.30\textwidth,angle=270]{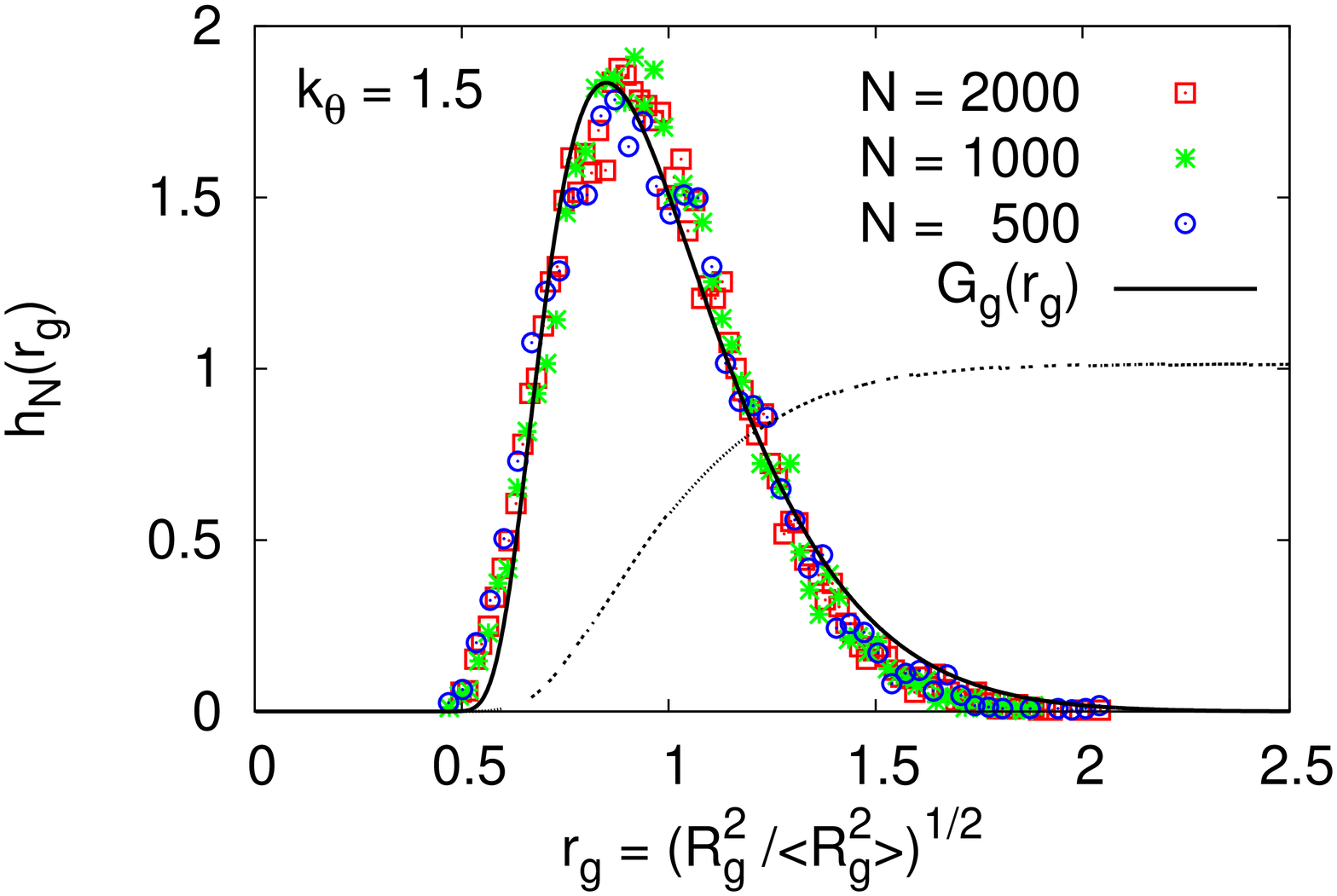}
\caption{Normalized probability distributions of
$r_e=(R_e^2/\langle R_e^2 \rangle)^{1/2}$, $h_N(r_e)$ (a), and
$r_g=(R_g^2/\langle R_g^2 \rangle)^{1/2}$, $h_N(r_g)$ (b), plotted versus
$r_e$, and $r_g$, respectively, for polymer chains in a melt. Data are for
$N=500$, $1000$, and $2000$. Theoretical predictions 
$G_e(r_e)$ \{Eq.~(\ref{eq-Ge})\}
and $G_g(r_g)$ \{Eq.~(\ref{eq-Gg})\} with $b_1=0.14$, $b_2=1.52$,
and $C_g=7.92$ such that $\int_0^\infty dr_g G_g(r_g)=1$ are
also shown in (a) and (b), respectively, for comparison.}
\label{fig-dist-ReRg2}
\end{center}
\end{figure}

  The probability distribution of any observable $x$ is normally obtained numerically by
accumulating the histogram $H_N(x)=\sum_{config.} \delta_{x,x'}$ 
over all configurations and all chains of
size $N$, and then normalizing the histogram such that  
\begin{equation}
       h_N(x)=\frac{H_N(x)}{\sum_{x'} H_N(x')} \quad \textrm{and} \quad
     \sum_x h_N(x)=1
\label{eq-h}
\end{equation}

In Fig.~\ref{fig-dist-ReRg2}, we present the normalized 
probability distribution $h_N(r_e)$ ($h_N(r_g)$) as a function of 
$r_e=(R_e^2/\langle R_e^2 \rangle)^{1/2}$ ($r_g=(R_g^2/\langle R_g^2 \rangle)^{1/2}$)
for polymer melts of three different chain sizes $N=500$, $1000$, $2000$. 
Note that an angular average over all directions
has been included in $h_N(r_e)$. 
We see the nice data collapse for both $h_N(r_e)$ and $h_N(r_g)$,
and they are described very well by the following two $N$-independent normalized 
distribution functions obtained from Eqs.~(\ref{eq-pRe}), (\ref{eq-pRg}),
and $\langle R_g^2 \rangle=k \ell_b^2 N$ with $k=0.4839$ shown in Fig.~\ref{fig-ReRg2},
\begin{eqnarray}
     G_e(r_e)=4 \pi r_e^2 \left(\frac{3}{2\pi} \right)^{3/2} 
\exp \left(-\frac{3r_e^2}{2} \right)  \, , \\
\qquad \int_0^\infty d r_e G_e(r_e)=1 \nonumber
\label{eq-Ge}
\end{eqnarray}
and
\begin{eqnarray}
     G_g(r_g)= C_g \exp
\left(-b_1 r_g^{-\alpha d} -b_2 r_g^\delta \right) \,, \\
\qquad \int_0^\infty d r_g G_g(r_g)=1 \nonumber
\label{eq-Gg}
\end{eqnarray}
where the parameters $b_1=a_1 k^{-\alpha d/2}$, $b_2=a_2 k^{\delta/2}$, and 
the normalization factor $C_g$ are determined numerically by 
a least-squares fit.

For understanding the connectivity
and correlation between monomers
the conformations of linear chains of contour length $L=(N-1)\ell_b$ in a melt
are usually described
by the average mean square internal distance,
$\langle R^2(s) \rangle$,
\begin{equation}
    \langle R^2(s) \rangle = \left \langle \frac{1}{n_c}
\sum_{i=1}^{n_c} \left[ \frac{1}{N-s} \sum_{j=1}^{N-s} (\vec{r}_{i,j}-\vec{r}_{i,j+s})^2 
\right] \right \rangle \, ,
\end{equation}
where $s$ is the chemical distance between the $j^{\rm th}$ monomer and
the $(j+s)^{\rm th}$ monomer along the identical chain.
It is generally believed that the theoretical prediction of mean square internal distance
for polymer melts consisting of semiflexible chains in
the absence of excluded volume effect described by a freely rotating chain model 
is~\cite{Flory1969}
\begin{equation}
   \langle R^2(s) \rangle =s\ell_b^2 \left[ 
\frac{1+\langle \cos \theta \rangle}{1- \langle \cos \theta \rangle}
-\frac{2 \langle\cos \theta \rangle( 1-\langle \cos \theta \rangle^s)}
{s(1-\langle \cos \theta \rangle)^2} \right]  \, ,
\label{eq-Rs2-FRC}
\end{equation}
with
\begin{equation}
 \langle \cos \theta \rangle = \langle \vec{b}_j \cdot \vec{b}_{j+1} \rangle /\ell_b^2
\,,\qquad j=1,2,\ldots,N-1 \,.
\end{equation}
In the limit $N\rightarrow \infty$,
the bond-bond orientational correlation function therefore
decays exponentially as a function of
chemical distance $s$ between any two bonds along a linear
chain~\cite{Grosberg1994,Rubinstein2003},
\begin{equation}
 \langle  \vec{b}_j \cdot \vec{b}_{j+s} \rangle =\ell_b^2 
\langle \cos \theta(s) \rangle =\ell_b^2 \langle \cos \theta \rangle^s
=\ell_b^2 \exp(-s \ell_b / \ell_p) \,,
\label{eq-cos-dWLC1}
\end{equation}
where $\ell_p$ is the so-called persistence length which can be
extracted from the initial decay of $\langle \cos \theta (s) \rangle$.

\begin{figure}[t]
\begin{center}
\includegraphics[width=0.30\textwidth,angle=270]{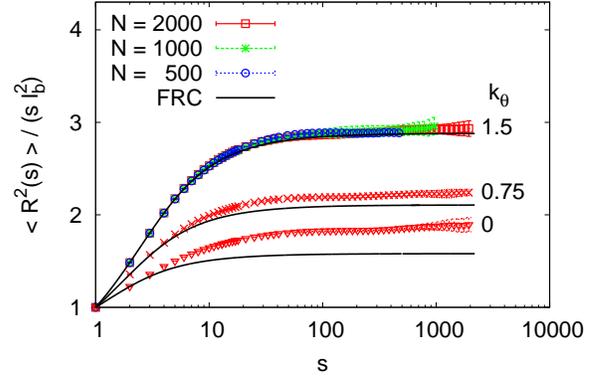} \hspace{0.4cm}
\caption{Rescaled mean square internal distance,
$\langle R^2(s) \rangle/(s \ell_b^2)$, plotted as a function
of $s$ with error bars.
Three different chain sizes $N=500$, $1000$, and $2000$ are chosen
for $k_\theta=1.5$, as indicated. For $k_\theta=0.75$ and $k_\theta=0$,
only data for $N=2000$ are included. The theoretical prediction for
freely rotating chains (FRC) is also shown by solid curves for comparison.
}
\label{fig-Rs2}
\end{center}
\end{figure}

As $s=N-1$, Eq.~(\ref{eq-Rs2-FRC})
gives the asymptotic behavior of the mean square end-to-end distance of
a FRC equivalent to the behavior of a freely jointed chain
\begin{eqnarray}
  \langle R_e^2 (N)\rangle &=& C_\infty (N-1)\ell_b^2 \, \thinspace {\rm with} \thinspace \,
\,  C_\infty=\frac{1+\langle \cos \theta \rangle}{1-\langle \cos \theta \rangle} \quad
\label{eq-Re2-N} \\
&=& \ell_K L= 2 \ell_p L
\label{eq-Re2-FRC}
\end{eqnarray}
where $C_\infty$ is so-called Flory's characteristic ratio~\cite{Flory1969},
and $\ell_K=2\ell_p$ is the Kuhn length.

\begin{figure}[t]
\begin{center}
(a)\includegraphics[width=0.30\textwidth,angle=270]{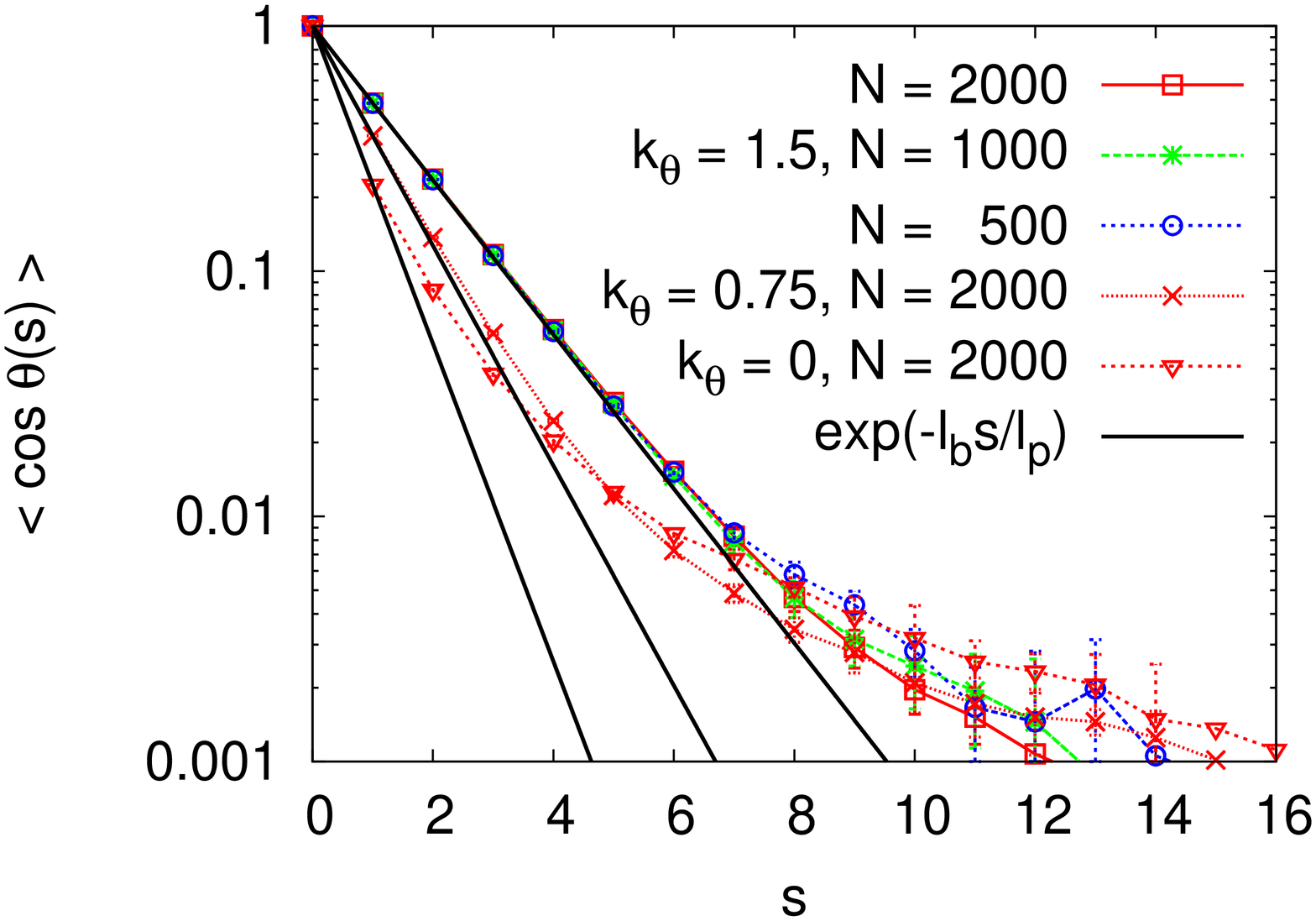} \hspace{0.4cm}
(b)\includegraphics[width=0.30\textwidth,angle=270]{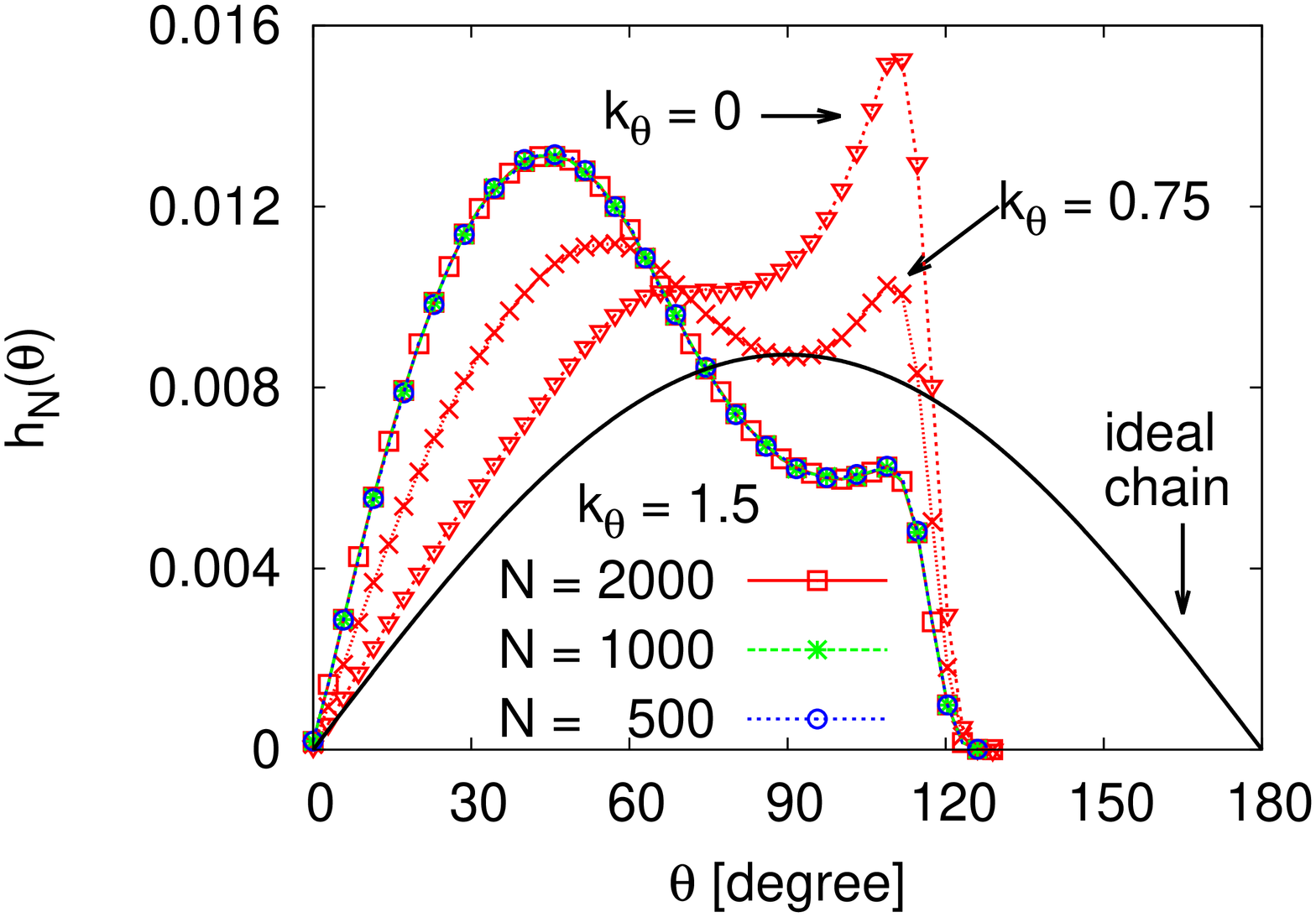}
\caption{
(a) Semi-log plot of the bond-bond orientational correlation function
$\langle \cos \theta(s) \rangle$ vs.~$s$ with error bars.
(b) Normalized probability distribution of bond angles $\theta$, $h_N(\theta)$,
plotted versus $\theta$. 
In (a), the straight lines indicate the initial exponential decay 
$\exp(-\ell_bs/\ell_{p})$ with $\ell_p/\ell_b=0.67$,
$0.97$, and $1.38$ for $k_\theta=0$, $0.75$, and $1.5$, respectively.
In (b), the theoretical prediction (solid curve) for an ideal chain
in a dilute solution is also shown for comparison.
Data are for polymer melts containing $n_c=1000$ chains of $N$ monomers.
$N=500$, $1000$, $2000$ for $k_\theta=1.5$, and $N=2000$
for $k_\theta=0.75$, $0$, as indicated.
}
\label{fig-ptheta}
\end{center}
\end{figure}

  Results of $\langle R^2(s) \rangle $ scaled by ($s \ell_b^2$), 
obtained by taking the average over ${\cal O}(10)$ independent 
polymer melts containing $1000$ chains to reduce fluctuations at large $s$, are shown in Fig.~\ref{fig-Rs2}.
For $k_\theta=1.5$, we see the nice data collapse for chains of different sizes $N$.
The universal scaling behavior for $k_\theta=1.5$ is nearly in perfect agreement with 
the theoretical prediction of $\langle R^2(s) \rangle$ for semiflexible chains in the 
absence of excluded volume effect described by a
freely rotating chain (FRC) model. However, a slight deviation from the predicted 
curve for FRC occurs for $N>800$. This deviation becomes more prominent 
as the flexibility of polymer chains increases due to the correlation hole effects
that the correlation hole is deeper for more flexible chains.
Note that here we do not take 
the bond-bond orientational correlation between two successive
bond vectors, $\langle \cos \theta \rangle$ in Eq.~(\ref{eq-Rs2-FRC}),
as a fitting parameter~\cite{Auhl2003}, but rather we estimate $\langle \cos \theta \rangle$
directly from the equilibrated configurations of polymer melts.

  The correlations $\langle \cos \theta(s) \rangle$ between two bonds along an identical chain 
at a chemical distance $s$ for $k_\theta=1.5$, $0.75$, $0$ are shown in Fig.~\ref{fig-ptheta}a.
As it was clarified in Refs.~\cite{Wittmer2007,Hsu2010,Hsu2014m}, the asymptotic
decay of $\langle \cos \theta(s) \rangle$ as a function of $s$ 
for dense melts and at the $\Theta$ point is not a single
exponential as predicted by Eq.~(\ref{eq-cos-dWLC1}), but rather a power law
decay, $\langle \theta(s) \rangle \propto s^{-3/2}$ for $s^* \ll s \ll N$,
due to excluded volume effects. 
Therefore only the initial decay of $\langle \cos \theta(s) \rangle$
is meaningful for the estimation of the persistence length $\ell_p$.
However, the crossover point $s^*$ shifts to
larger value of $s$ as the chain stiffness increases, i.e. 
the range over which the exponential decay holds extends.
We also check how the profiles of probability distribution $P(\theta)$
of bond angles $\theta$ vary with increasing chain stiffness.
Using Eq.~(\ref{eq-h}), $P(\theta)$ is estimated by
accumulating normalized histograms $h_N(\theta)$ of $\theta$ between two successive bonds along a chain.
We see that in Fig.~\ref{fig-ptheta}b, the distributions have a bimodal form.
For fully flexible chains ($k_\theta=0$) in a melt, 
there exists one peak occurring at $\theta \approx 110^o$ 
due to the competition between the excluded volume effect and the flexibility. 
As the chain stiffness increases 
($k_\theta$ increases), a second peak starts to develop at $\theta<90^o$, and
the position where the peak is located shifts to a smaller value of $\theta$.
For an ideal chain in a dilute solution, one should expect that 
\begin{equation}
   P(\theta)=\frac{1}{2}\sin \theta \,, \qquad \int_0^{\pi} d\theta P(\theta)=1 \,.
\label{eq-ptheta}
\end{equation}
This is also shown in Fig.~\ref{fig-ptheta}b for comparison.

\begin{figure}[t]
\begin{center}
(a)\includegraphics[width=0.30\textwidth,angle=270]{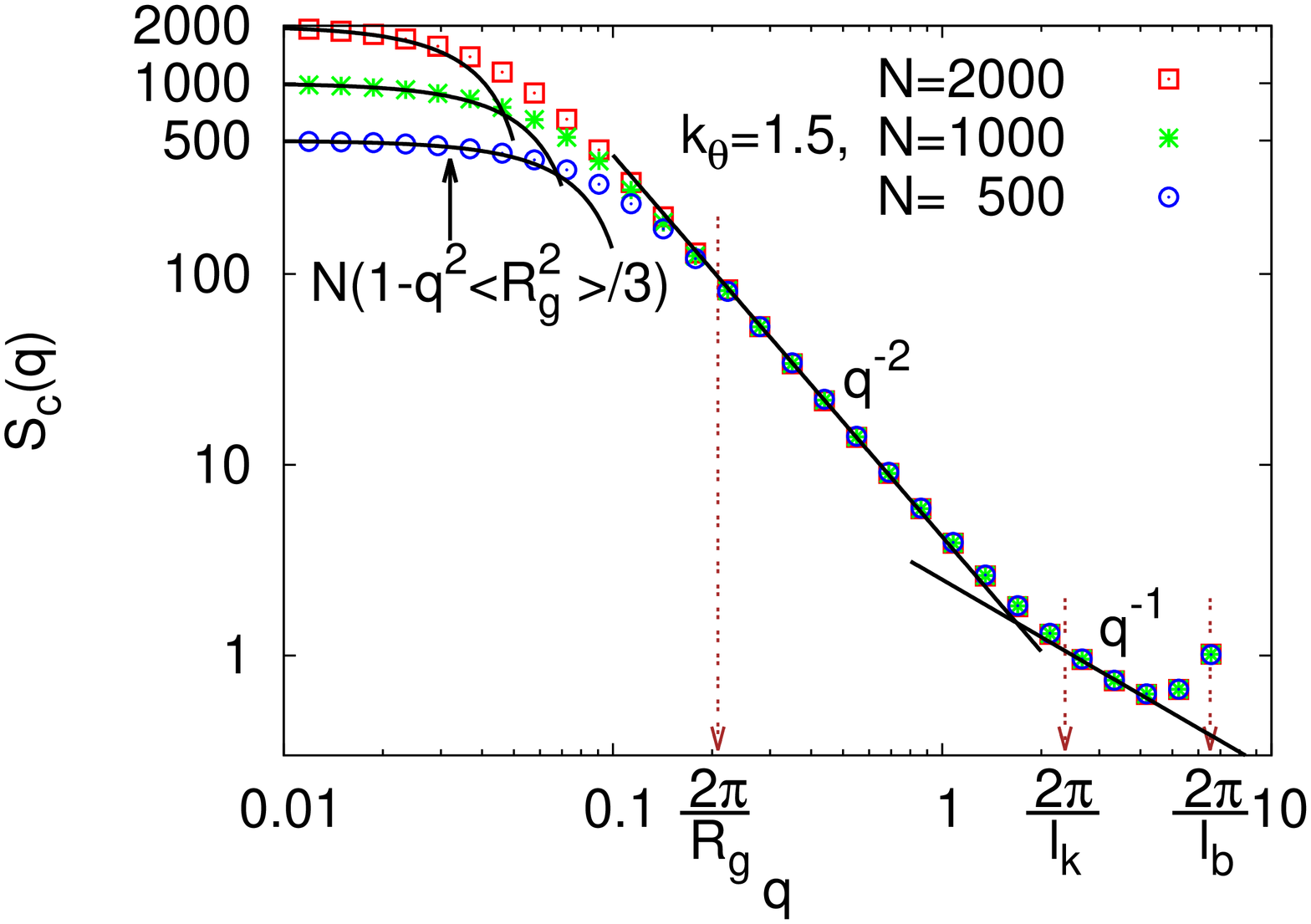}\hspace{0.4cm}
(b)\includegraphics[width=0.30\textwidth,angle=270]{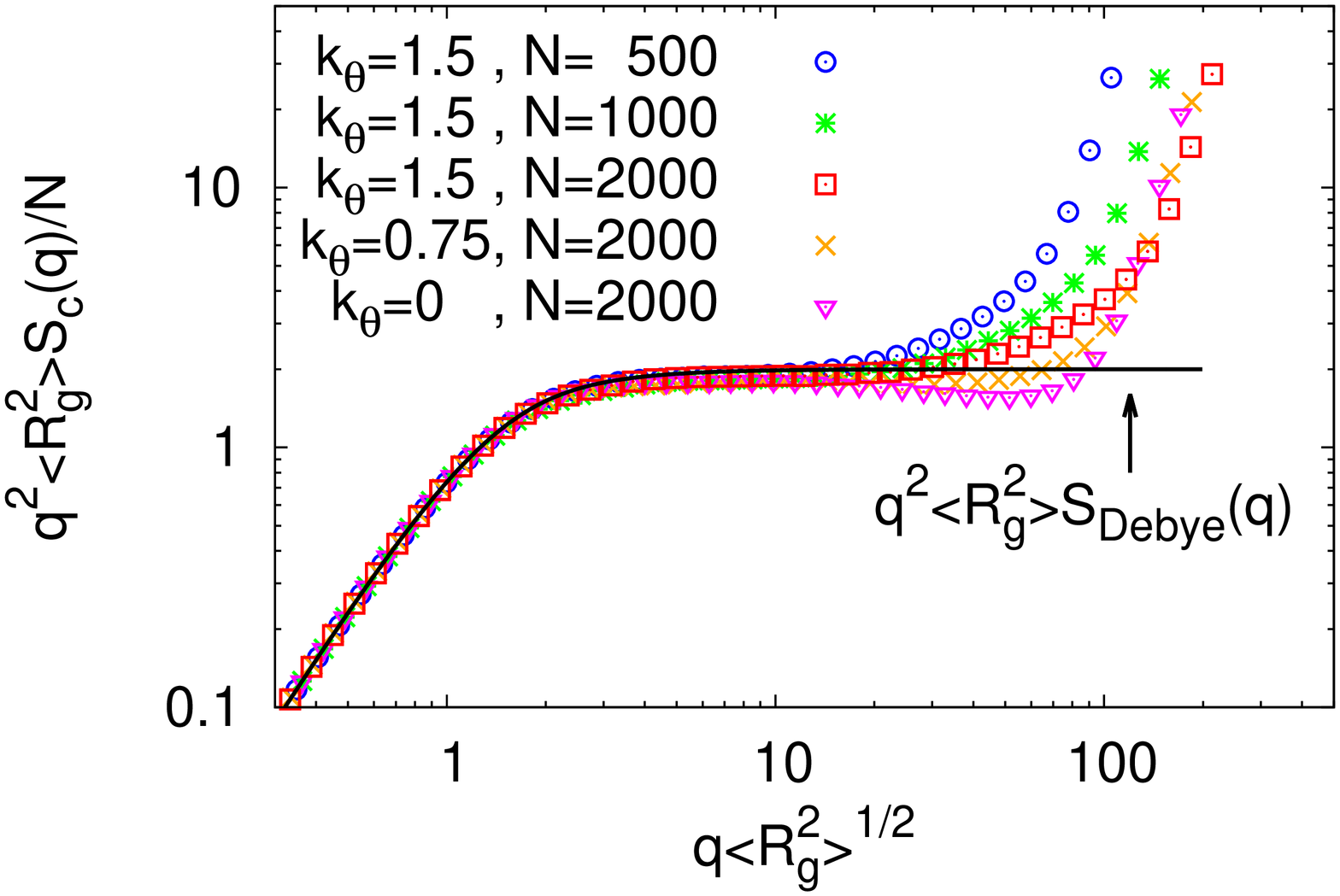}
\caption{(a) Structure factors of single chains in a melt, $S_c(q)$,
plotted vs.\ $q$ on log-log scales
for polymer melts consisting of $n_c=1000$ chains of $N=500$, $1000$,
and $2000$ monomers, and for $k_\theta=1.5$, as indicated.
(b) Same data as in (a) but in a Kratky-Plot. Data for $N=2000$, and
$k_\theta=0$ and $0.75$ are also shown, as indicated.
In (a) the theoretical predictions
$S_c(q)=N(1-q^2 \langle R_g^2 \rangle/3)$ at the Guinier regime for small $q$,
$S_c(q)\sim q^{-2}$ for a Gaussian coil, and $S_c(q) \sim q^{-1}$ for a rigid rod
are shown by solid curves for comparison. The theoretically predicted
cross-over points between different regimes are pointed out by arrows.
Here the root-mean-square gyration radius 
$R_g=\langle R_g^2(N=2000) \rangle^{1/2} \approx 30.15$ 
($R_e \approx 73.44$), the Kuhn length 
$\ell_k \approx  2.66$, and the root-mean-square bond length $\ell_b \approx 0.964$.
In (b) the Debye function, Eq.~(\ref{Debye}), are
also shown by a solid curve for comparison.}
\label{fig-sqc}
\end{center}
\end{figure}

The scattering from single chains in a melt in equilibrium
is shown in Fig.~\ref{fig-sqc}.
In Fig.~\ref{fig-sqc}a we see that
$S_c(q) \approx N\exp(-q^2 \langle R_g^2 \rangle/3)
\approx N(1-q^2 \langle R_g^2 \rangle/3)$ for small $q$ 
($q \ll \frac{2\pi}{R_g}$, $R_g=\sqrt{\langle R_g^2 \rangle}$)
in the Guinier regime, then a
crossover occurs to the power law of Gaussian coils (ideal chains),
$S(q) \sim q^{-1/\nu}$ with $\nu=1/2$ 
for $\frac{2\pi}{R_g}<q<\frac{2\pi}{\ell_k}$. Here 
$\ell_k=2\ell_p\approx 2.66$ for $k_\theta=1.5$
using Eq.~(\ref{eq-cos-dWLC1}).
Though our chains are moderately stiff ($k_\theta=1.5$) the short range initial rigid-rod regime
$S(q) \sim q^{-1}$ for $2\pi/\ell_k < q \ll 2\pi/\ell_b$ is hardly visible, thus allowing them
still to be taken as a model for flexible polymers. 
In order to clarify whether single chains in a melt behave as ideal chains
we show the structure factors $S_c(q)$ in a Kratky-plot in Fig.~\ref{fig-sqc}b.
The Debye function~\cite{deGennes1979,Cloizeaux1990,Schaefer1999,Higgins1994}
describing the scattering from Gaussian chains,
\begin{equation}
 S_{\rm Debye}(q)=2\frac{\eta-1+\exp(-\eta)}{\eta^2} \quad {\rm with} \quad 
\eta=q^2 \langle R_g^2 \rangle \,,
\label{Debye}
\end{equation}
is also presented in Fig.~\ref{fig-sqc}b for comparison. 
The deviations from ideality are clearly recognized near
$q\langle R_g^2 \rangle^{1/2} \approx 20$ 
for rather flexible chains ($k_\theta=0$, $k_\theta=0.75$) of size $N=2000$,
and a minimum value is reached in the Kratky-plot as $q$ increases,
in agreement with the previous work~\cite{Wittmer2007,Hsu2014m}.
As a first conclusion one can state that polymer melts of chains with 
a stiffness parameter $k_\theta=1.5$ offer a good compromise for modeling 
highly flexible polymers while at the same time minimizing deviations from
ideality, which significantly impair the use of simple models for fully
flexible chains.

\section{Dynamic properties of equilibrated polymer melts}
\label{Dynamics}

The dynamic behavior of polymer chains in a melt or solution 
is usually characterized by the mean square displacement (MSD) of monomers.
The theoretical
predictions of the dynamic scaling behavior of MSD given by
the reptation theory~\cite{deGennes1979, Doi1986}
show that the crossover behavior occurs at different time scales,
the characteristic time $\tau_0$, the entanglement time $\tau_e \sim \tau_0 N_e^2$,
the Rouse time $\tau_R \sim  \tau_0 N^2$, and the disentanglement
time $\tau_d \sim \tau_0 N^3/N_e$ (in the ideal case where
the chain length $N$ is very large). 
However, all simulations and experiments 
support $\tau_d \propto N^{3.4}$ due to the reason that contour length
fluctuation, constrains release and correlation hole effects 
shift the crossover to the asymptotic behaviors to very long 
chains~\cite{Doi1983,Milner1998,McLeish2002,Likhtman2002}.

   Three quantities describing the dynamic properties of polymer chains in a melt
are listed as follows:
the mean square displacement of a monomer,
\begin{eqnarray}
g_1(t) \equiv \frac{1}{n_c(\frac{N}{2}+1)} \sum_{i=1}^{n_c}\sum_{j=N/4}^{3N/4} \left \langle \left[
\vec{r}_{i,j}(t)-\vec{r}_{i,j}(0) \right]^2 \right \rangle \; ,
\label{eq-g1}
\end{eqnarray}
the mean square 
displacement of monomers with respect to the corresponding center of mass (c.m.),
\begin{eqnarray}
&g_2(t)& \equiv \frac{1}{n_cN} \sum_{i=1}^{n_c}\sum_{j=1}^{N}  \nonumber \\
&&\left \langle \left[
(\vec{r}_{i,j}(t)-\vec{r}_{i,{\rm c.m.}}(t))  
-(\vec{r}_{i,j}(0)-\vec{r}_{i,{\rm c.m.}}(0))
 \right]^2 \right \rangle \,, \nonumber \\
\label{eq-g2}
\end{eqnarray}
and the mean square displacement of the center of mass
\begin{eqnarray}
g_3(t) \equiv \left \langle \left[
\vec{r}_{i,{\rm c.m.}}(t)-\vec{r}_{i,{\rm c.m.}}(0) \right]^2 \right \rangle \,, \nonumber\\
\vec{r}_{i,{\rm c.m.}}(t)=\frac{1}{N}\sum_{j=1}^N \vec{r}_{i,j}(t) \,. 
\label{eq-g3}
\end{eqnarray}
Note that in Eq.~(\ref{eq-g1}) only half of the monomers in 
the middle of each chain are considered in order to suppress the fluctuations
caused by chain ends~\cite{Kremer1990,Kremer1992}, while all monomers in
each chain $i$ are considered in the calculation of the center of mass
$\vec{r}_{i,{\rm c.m.}}(t)$ \{Eq.~(\ref{eq-g3})\}.
The corresponding scaling predictions of $g_1(t)$, $g_2(t)$,
and $g_3(t)$ are given by~\cite{Kremer1990,Puetz2000} 
\begin{widetext}
\begin{eqnarray}
\begin{array}{lll}
 g_1(t) \sim \left \{ \begin{array}{lc}
  t^1 \;,  &  t<\tau_0 \\
  t^{1/2}\;,  & \tau_0<t<\tau_e  \\
  t^{1/4}\;,  & \tau_e<t<\tau_R  \\
  t^{1/2}\;, & \tau_R<t<\tau_d  \\
  t^1\; ,  & t >\tau_d 
 \end{array} \right . \,,
& 
g_2(t) \sim \left \{ \begin{array}{lc}
  t^1 \;, &  t<\tau_0 \\
 t^{1/2} \;, & \tau_0<t<\tau_e \\
  t^{1/4}\;, & \tau_e<t<\tau_R \\
  t^{1/2}\;, & \tau_R<t<\tau_d \\
  t^{0}\;, & t>\tau_d  
 \end{array} \right . \, ,
&
  g_3(t) \sim \left \{ \begin{array}{lc}
   t^1 \;, &  t<\tau_e \\
   t^{1/2} \;,  & \tau_e<t<\tau_R \\
   t^1\;,  & t >\tau_R 
\end{array} \right . \, . 
\end{array} \nonumber
\\
\label{eq-g123-s}
\end{eqnarray}
\end{widetext}

\begin{figure*}[!thb]
\begin{center}
(a)\includegraphics[width=0.30\textwidth,angle=270]{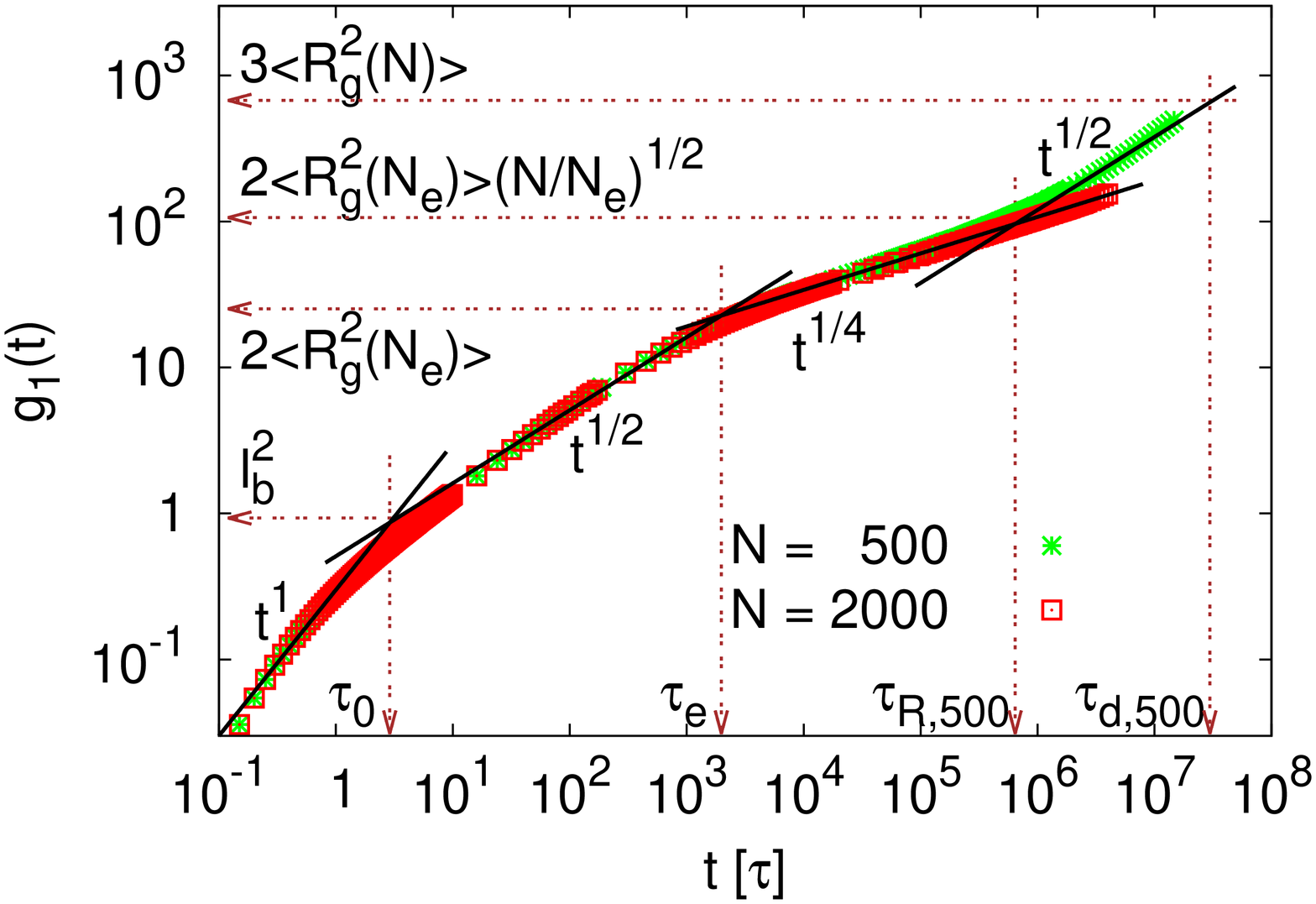}\hspace{0.4cm}
(b)\includegraphics[width=0.30\textwidth,angle=270]{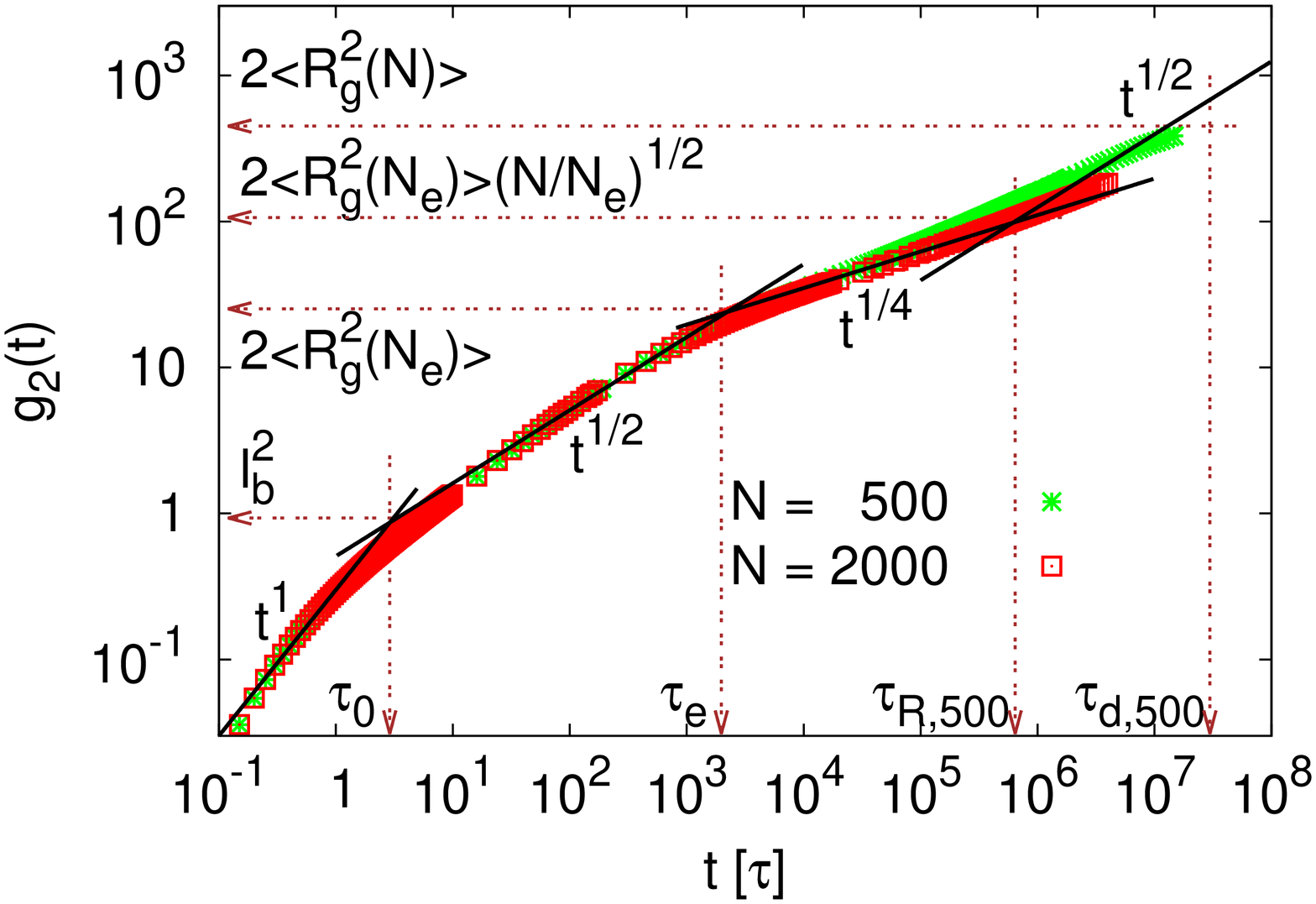}\\
(c)\includegraphics[width=0.30\textwidth,angle=270]{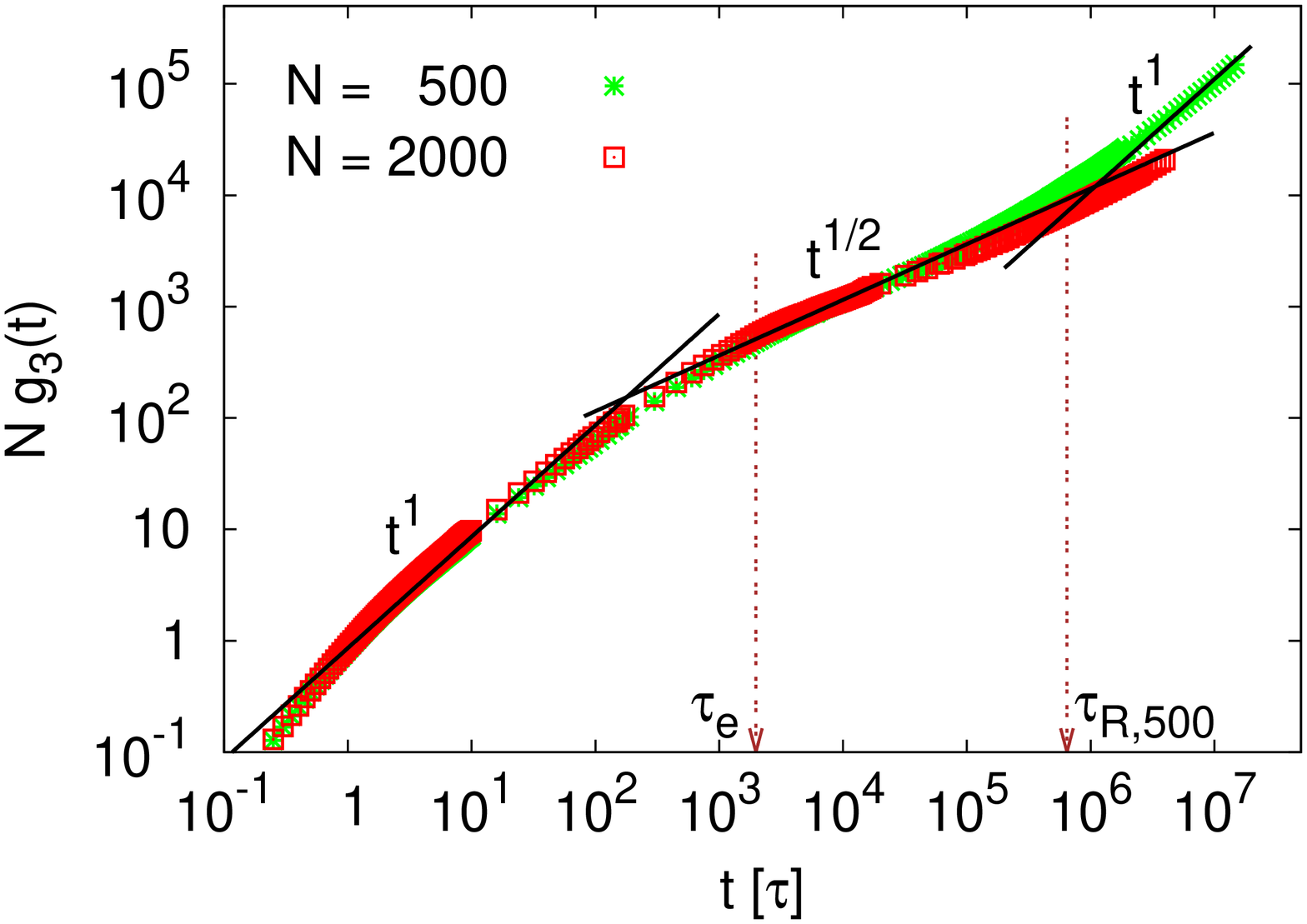}\hspace{0.4cm}
(d)\includegraphics[width=0.30\textwidth,angle=270]{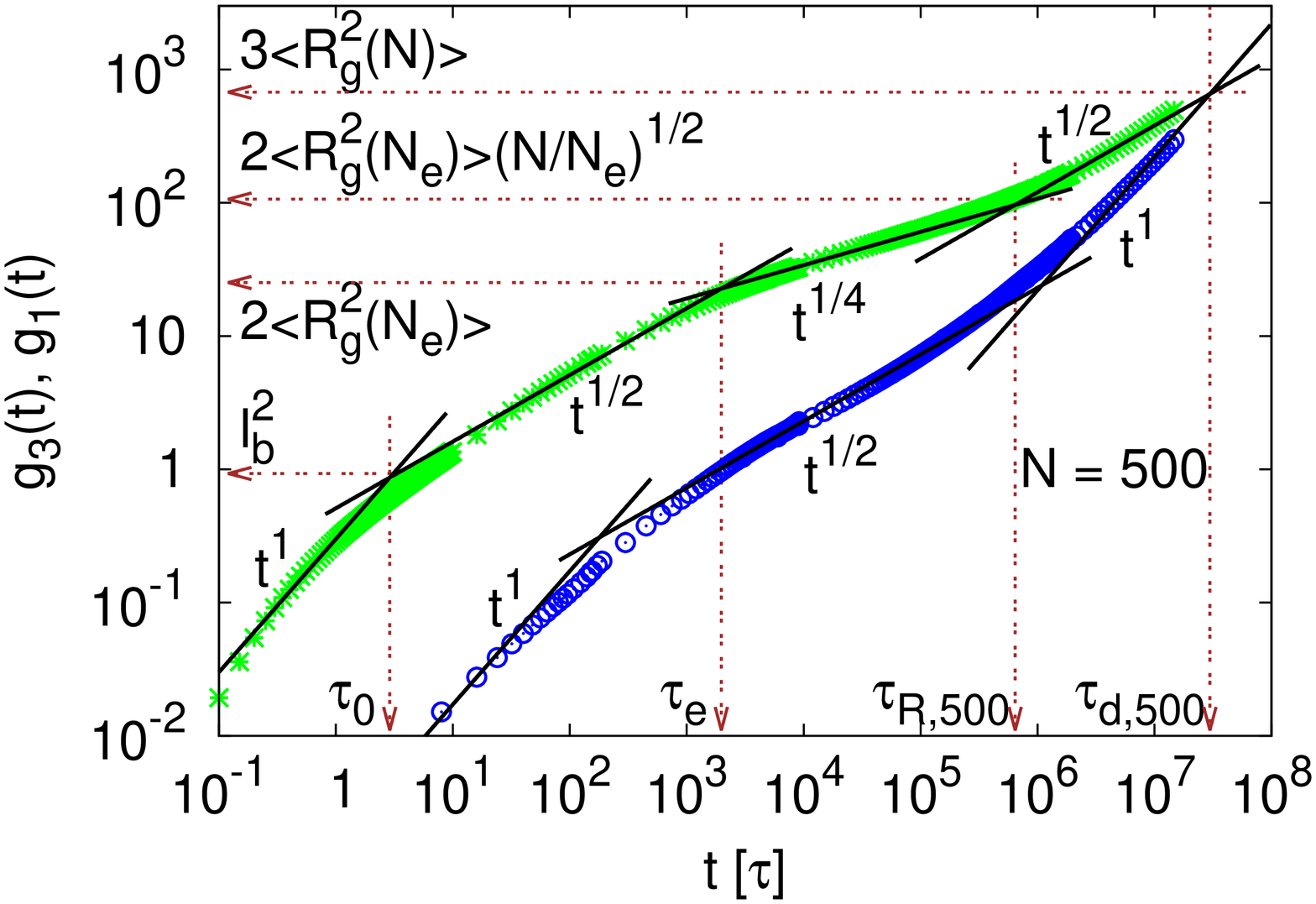}
\caption{Mean square displacement of inner monomers $g_1(t)$ (a), monomers
with respect to the center of mass of the corresponding chain, $g_2(t)$ (b),
and center of mass, and $g_3(t)$ multiplied by $N$ (c), plotted versus $t$ for
$N=500$ and $N=2000$, as indicated. (d) $g_1(t)$ and $g_3(t)$ versus $t$ for $N=500$.
The crossover points between two different scaling
regimes are determined by the intersections of two straight lines shown in (a), and
marked by arrows at $t=\tau_0 \approx 2.89 \tau$, $\tau_e \approx 1.98\times 10^3 \tau$,
and $\tau_{R,500} \approx 6.44\times 10^5 \tau$ 
while $\tau_{d,500} \approx 2.97 \times 10^7 \tau$ is
determined by the intersection of the two neighboring fitting curves for $g_1(t)$ and $g_3(t)$ at
$t\gg \tau_{R,500}$ in (d).}
\label{fig-g123}
\end{center}
\end{figure*}

  Our extensive molecular dynamics results of $g_1(t)$, $g_2(t)$, and $g_3(t)$ up to 
$t \sim {\cal O}(10^7) \tau$  
for polymer chains of sizes $N=500$, $2000$ in a melt are shown in Fig.~\ref{fig-g123}. The best fits of the theoretical 
predictions given in Eq.~(\ref{eq-g123-s}) are shown by
solid lines for comparison. 
The characteristic time scales $\tau_0 \approx 2.89 \tau$ where $\tau$ is the LJ
time unit (see Appendix), 
$\tau_e=\tau_0 N_e^2 \approx 1.98 \times 10^3 \tau$, 
and $\tau_{R,500}=\tau_0 N^2 \approx 6.44 \times 10^5 \tau$ for $N=500$
are determined by the intersection points of 
two lines from the results of $g_1(t)$ in Fig.~\ref{fig-g123}a.
They correspond to the crossover points between two scaling regimes are pointed out
by arrows also in Fig.~\ref{fig-g123}bcd.
The disentanglement time 
$\tau_d \approx 2.97 \times 10^7 \tau$ is determined 
from the intersection
between the fitting straight lines of $g_1(t) \propto t^{1/2}$ for $\tau_d>t>\tau_R$
and $g_3(t) \propto t^1$ for $t>\tau_R$, respectively, since
we should expect that $g_1(t)=g_3(t)$ for $t>\tau_d$. 
The characteristic time $\tau_0$ estimated from $\tau_{R}=\tau_0N^2$ for 
$N=500$ is $2.58 \tau$ which is compatible with the direct measurement.
If we estimate the entanglement length $N_e$ from characteristic time scales 
$\tau_0$, $\tau_e$, $\tau_R$ and $\tau_d$ determined by the scaling predictions of the
mean square displacement for $N=500$ (Fig.~\ref{fig-g123}a and \ref{fig-g123}d), 
we get $N_e = (\tau_e/\tau_0)^{1/2} \approx 26(1)$  
and $N_e=N(\tau_R/\tau_d)^{1/1.4} \approx 32(2)$ if we assume that
$\tau_d = \tau_R (N/N_e)^{1.4}$.
Both estimates are consistent with results from PPA and from the relaxation
plateau modulus within error bars (see Table~\ref{table1}).
The two estimates are deviating by about $10\%$ 
from the expected value $N_e = 28$. If we fit our data with $\tau_d=(N/N_e)\tau_R$,
we get $N_e \sim 11$ which is underestimate. Thus our data perfectly fit experiments
that $\tau_d \propto N^{3.4}$
and show the limitations of the asymptotic theory.

According to the theoretical predictions,
we see that in Fig.~\ref{fig-g123}a, $g_1(t) \approx \ell_b^2$ at $t=\tau_0$.
At $t=\tau_e$, 
$g_1(t) \sim d_T^2 \approx 2 \langle R_g^2(N_e) \rangle \approx (5.02 \sigma)^2$
(assuming that a Rouse chain of $N_e$ monomers is relaxed)~\cite{Kremer1990,Kremer1992},
where the entanglement effect starts to set in and monomers in an identical
chain are restricted
to move only along the contour of an imaginary tube of diameter $d_T$
and contour length 
$L_T=d_T(N/N_e)$ until reaching $t=\tau_{R,500}$ for $N=500$. Since the tube itself is a
random walk with a step length $d_T$, the displacement of a monomer at $t=\tau_{R,500}$ is thus
$g_1(t) \sim  d_T^2(N/N_e)^{1/2} \approx 2\langle R_g^2(N_e) \rangle (N/N_e)^{1/2}$.
In the case of $N=2000$, we find that our data of $g_1(t)$ 
($g_2(t)$) follow the power law $t^{1/4}$ about three decades for $t>\tau_e$, 
a much longer time window than observed so far via simulation.
For $\tau_d > t > \tau_R$, the polymer chain slides back
and forth along the tube-like regime and results in a second $g_1(t) \propto t^{1/2}$
regime which is predicted by the reptation theory~\cite{Doi1986}.
After reaching the disentanglement time (reptation time) $\tau_{d,500}$,
a chain has moved a distance comparable to its own size 
$g_1(t)=g_3(t) \approx 3 \langle R_g^2 (N)\rangle$ for $N=500$ (see Fig.~\ref{fig-g123}d).
The initial tube is completely destroyed and another new tube-like regime
will appear depending on the polymer chain size or
polymer molecular weight. Finally monomers diffuse such that $g_1(t) \propto t^1$
for $t>\tau_d$.

  Results of $g_1(t)$, $g_2(t)$, and $Ng_3(t)$ (Fig.~\ref{fig-g123}abc) show 
that they are all independent of $N$ for 
$t<\tau_{R,500}$. Furthermore $g_1(t) \cong g_2(t)$ in that regime. 
For $\tau_R< t < \tau_d$, either the size $N=500$ is still 
too short or the statistics for long relaxation time are insufficient,  
the expected scaling law $g_2(t) \propto t^{1/2}$ is only seen slightly,
while for $t \approx \tau_d$, $g_2(t)=\langle R_g^2(N) \rangle$ for
$N=500$ is barely reached.
However, such a proof for $N=2000$ or even longer chain lengths might only be possible
with further improved soft and hardware~\cite{Anderson2008, Glaser2015}.

\section{Comparison between the original chain conformations and the primitive path}
\label{PPA}

   In order to understand the structural differences between the original path
and the primitive path (pp) of polymer chains in a melt, we implement the same 
primitive path analysis
proposed by Everaers et al.~\cite{Everaers2004} based on the concept of
Edwards' tube model~\cite{Edwards1967} to identify the primitive path
of each polymer chain in a 
melt~\cite{Sukumaran2005, Kroeger2005, Shanbhag2005, Tzoumanekas2006,
Hoy2009, Everaers2012}. A detailed discussion regarding to self-entanglements,
local self-knot effect, and finite-size effect is given in Ref.~\cite{Sukumaran2005,Moreira2015}.

Since the motion of a chain is confined in a tube-like 
regime with fluctuation due to entanglements with other chains (see Sec.~\ref{Dynamics}), 
the primitive path of the chain is the contracted contour of an imaginary tube without
any other chain crossing when all endpoints are fixed in space.
In this analysis, topologies of chains are kept and 
chains are assumed to behave as random walks along their primitive paths. 
The mean square end-to-end distance of chains therefore remains the same as  
that for the original paths of chains, 
i.e., $\langle R_{e,pp}^2 \rangle=\langle R_{e}^2  \rangle$, and
\begin{eqnarray}
  \langle R_e^2(N) \rangle = 
\ell_{K}^{(pp)} L_{pp}=\ell_{K}^{(pp)}(N-1)l_b^{(pp)} \nonumber \\
{\rm with} \qquad 
\ell_b^{(pp)}=\frac{\sum_{j=1}^{N-1} \mid \vec{r}_{j+1}-\vec{r}_j \mid}{N-1} \,.
\label{eq-bpp}
\end{eqnarray} 
Here $\ell_{K}^{(pp)}$ is the Kuhn length, $L_{pp}$ is the contour length, 
and $\ell_b^{(pp)}$ is the average bond length of the primitive path. 
The so-called entanglement length 
$N_{e,PPA}$ 
defined by the number of monomers per Kuhn segment of the primitive path is then
\begin{equation}
       N_{e,PPA}=\frac{\ell_{K}^{(pp)}}{\ell_b^{(pp)}} \,.
\label{eq-Ne}
\end{equation}

\begin{figure}[t]
\begin{center}
(a)\includegraphics[width=0.30\textwidth,angle=270]{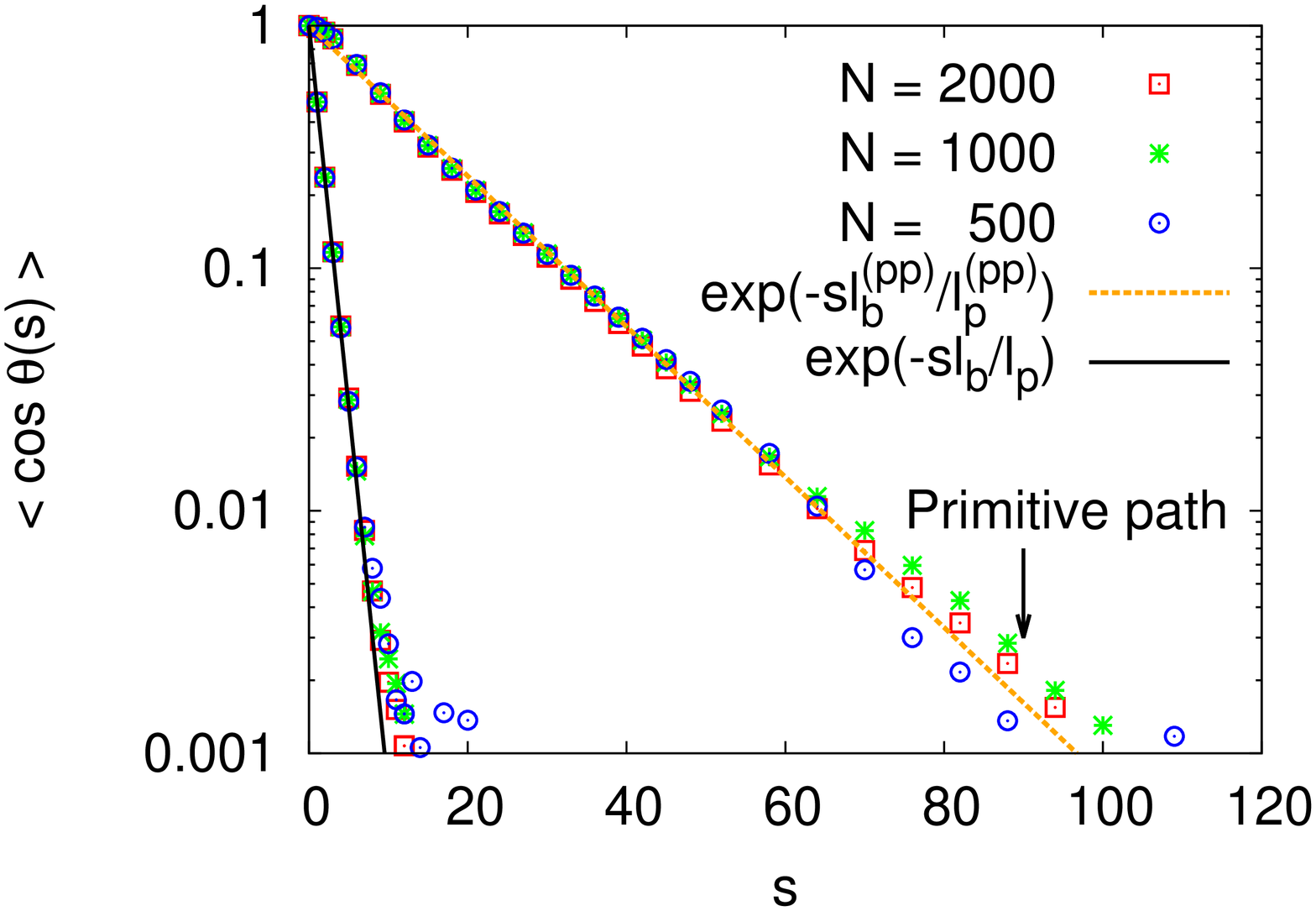} \hspace{0.4cm}
(b)\includegraphics[width=0.30\textwidth,angle=270]{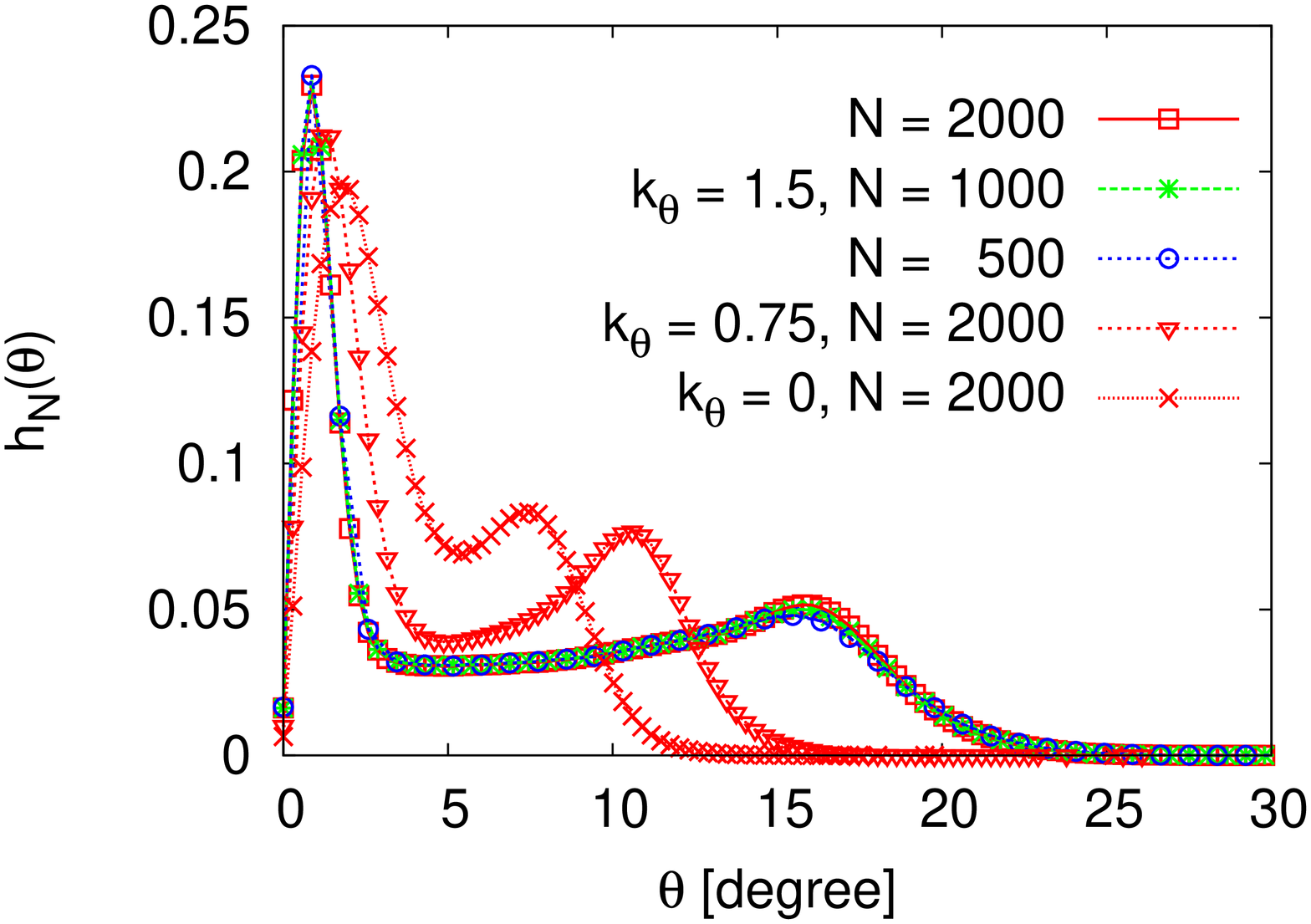}
\caption{
(a) Semi-log plot of the bond-bond orientational correlation function
$\langle \cos \theta(s) \rangle$ vs.~$s$.
(b) Normalized probability distribution of bond angles $\theta$, $h_N(\theta)$,
plotted versus $\theta$.
Data are for the primitive paths of polymer melts containing $n_c=1000$ chains of $N$ monomers.
$N=500$, $1000$, $2000$ for $k_\theta=1.5$, as indicated.
In (a), data for the original paths are also shown for comparison. The straight lines
indicate the initial exponential decay.
In (b), data for the primitive paths for $N=2000$, and for $K_\theta=0.75$,
$0$ are also shown for checking the effect of chain stiffness.}
\label{fig-ppa-theta}
\end{center}
\end{figure}

   Quantitatively, the primitive paths of all polymer chains in a melt are determined by
slowly cooling the system toward $T=0$ and minimizing the energy of the 
system~\cite{Everaers2004,Moreira2015}.
In the simulation, two ends of chains are fixed and the intrachain excluded volume 
interactions as well as the bond bending interactions are switched off while the 
interchain interactions are kept. 
In the case where the intrachain excluded volume
is kept, Sukumaran et. al.~\cite{Sukumaran2005} have found that the difference of 
the estimate of $N_{e,PPA}$ between these two cases is within error bars.
Results of the bond-bond orientational correlation function $\langle \cos \theta(s) \rangle$,
and the normalized histogram of bond angles $\theta$, $h_N(\theta)$ for 
the primitive paths of polymer chains in a melt with $k_\theta=1.5$ 
are shown in Fig.~\ref{fig-ppa-theta}. 
The initial decay of $\langle \cos \theta (s) \rangle$ 
described by an exponential decay $\exp(-s\ell_{b}^{(pp)}/\ell_{p}^{(pp)})$ 
up to $s=80$ 
is shown by a dashed line with 
$\ell_{p}^{(pp)}=\ell_K^{(pp)}/2$.
Since the endpoints of chains are fixed, without considering the interchain interactions 
and thermal fluctuations, chains are stretched out when the bond springs try to reduce 
the average bond length from $\ell_b=0.964$ to $\ell_{b}^{(pp)}=0.31$.
This effect is stronger at the short length scale ($s<10$) where the 
result of $\langle \cos \theta \rangle$
show some deviations from the fitting curve if we take a closer look.
The stretching conformations of chains are also observed from the 
normalized histogram $h_N(\theta)$ of bond angles $\theta$ shown
in Fig.~\ref{fig-ppa-theta}. The distribution of $\theta$ still has a 
bimodal form, but the range of 
$\theta$ shrinks from $[0^o,130^o]$ (Fig.~\ref{fig-ptheta}b) for the
original paths to $[0^o,30^o]$ for the primitive paths in the case of 
$k_\theta=1.5$. The distance between two peaks decreases as $k_\theta$
decreases.

\begin{figure}[t]
\begin{center}
\includegraphics[width=0.30\textwidth,angle=270]{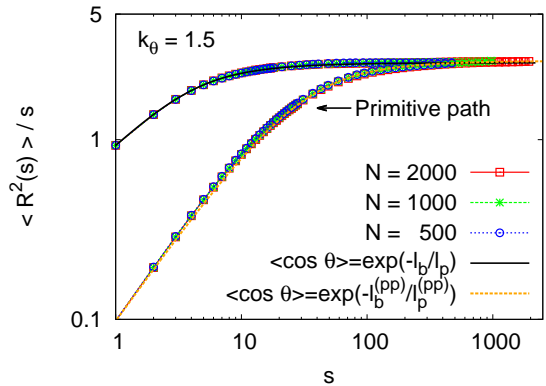}
\caption{Rescaled mean square internal distance,
$\langle R^2(s) \rangle/s$, plotted as a function
of $s$.
Data are for the original paths and the primitive paths of polymer
melts containing $n_c=1000$ chains of 
$N=500$, $1000$, $2000$ monomers, as indicated, and for $k_\theta=1.5$.
The theoretical predictions for
freely rotating chains (FRC) with $\langle \cos \theta \rangle = \exp(-s\ell_b/\ell_{p})$
for the original and the primitive paths, and
$\langle \cos \theta \rangle=\exp(-s\ell_{b}^{(pp)}/\ell_{p}^{(pp)})$  
with $\ell_{p}^{(pp)}=N_{e,PPA}\ell_b^{(pp)}/2$
for the primitive path are also shown for comparison.}
\label{fig-ppa-Rs2}
\end{center}
\end{figure}

\begin{figure*}[t]
\begin{center}
(a)\includegraphics[width=0.30\textwidth,angle=270]{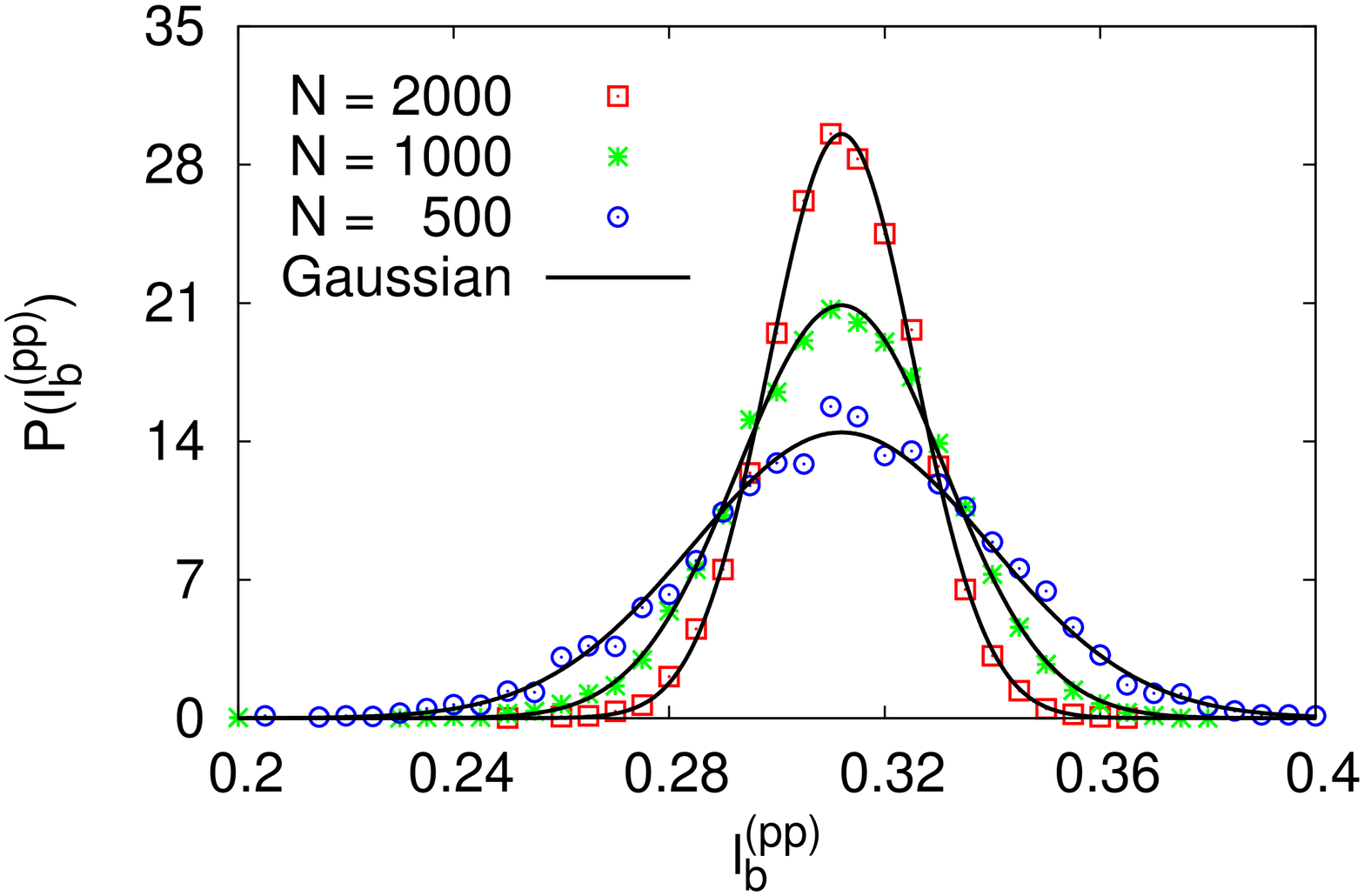} \hspace{0.4cm}
(b)\includegraphics[width=0.30\textwidth,angle=270]{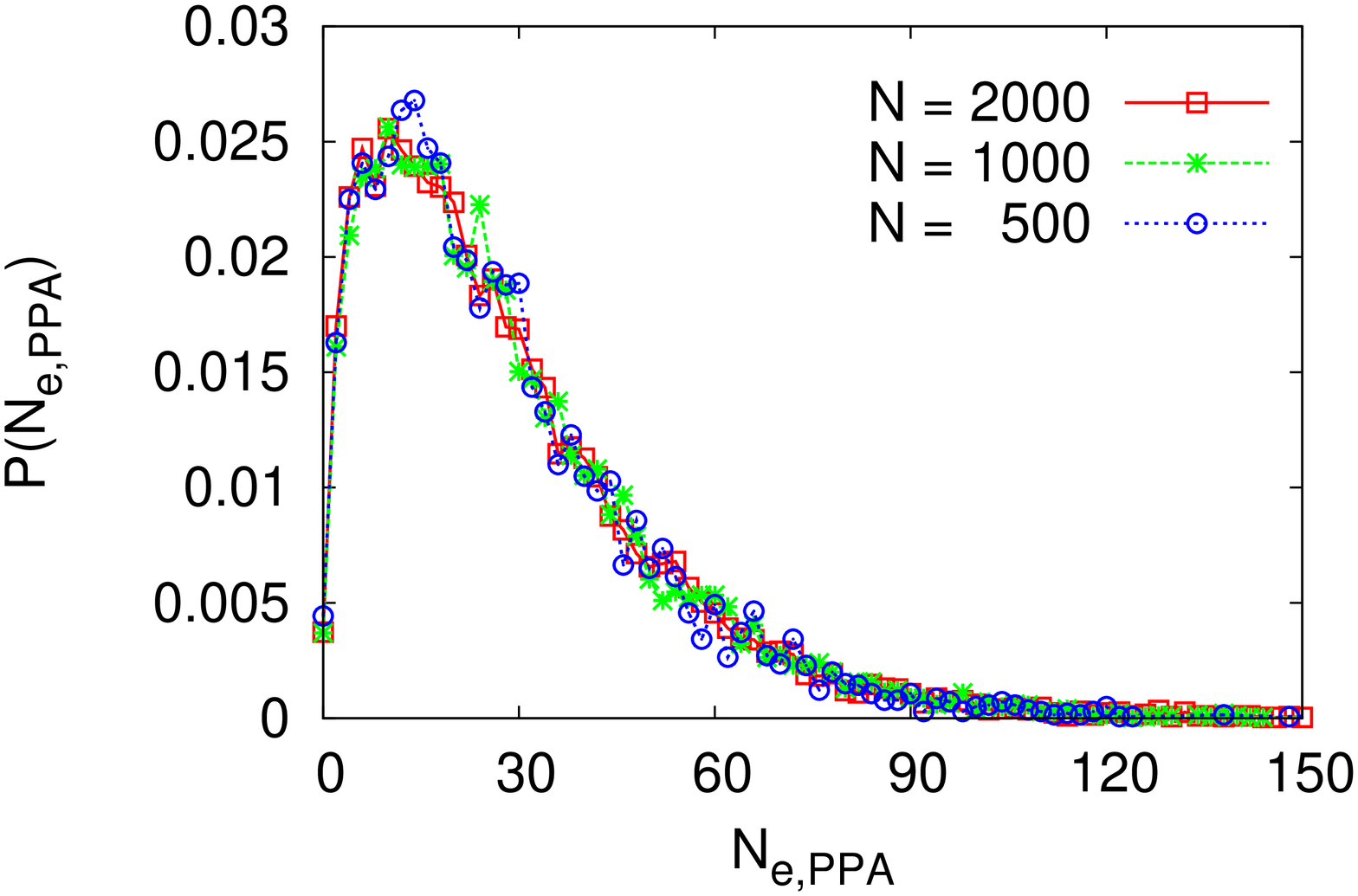}\\
(c)\includegraphics[width=0.30\textwidth,angle=270]{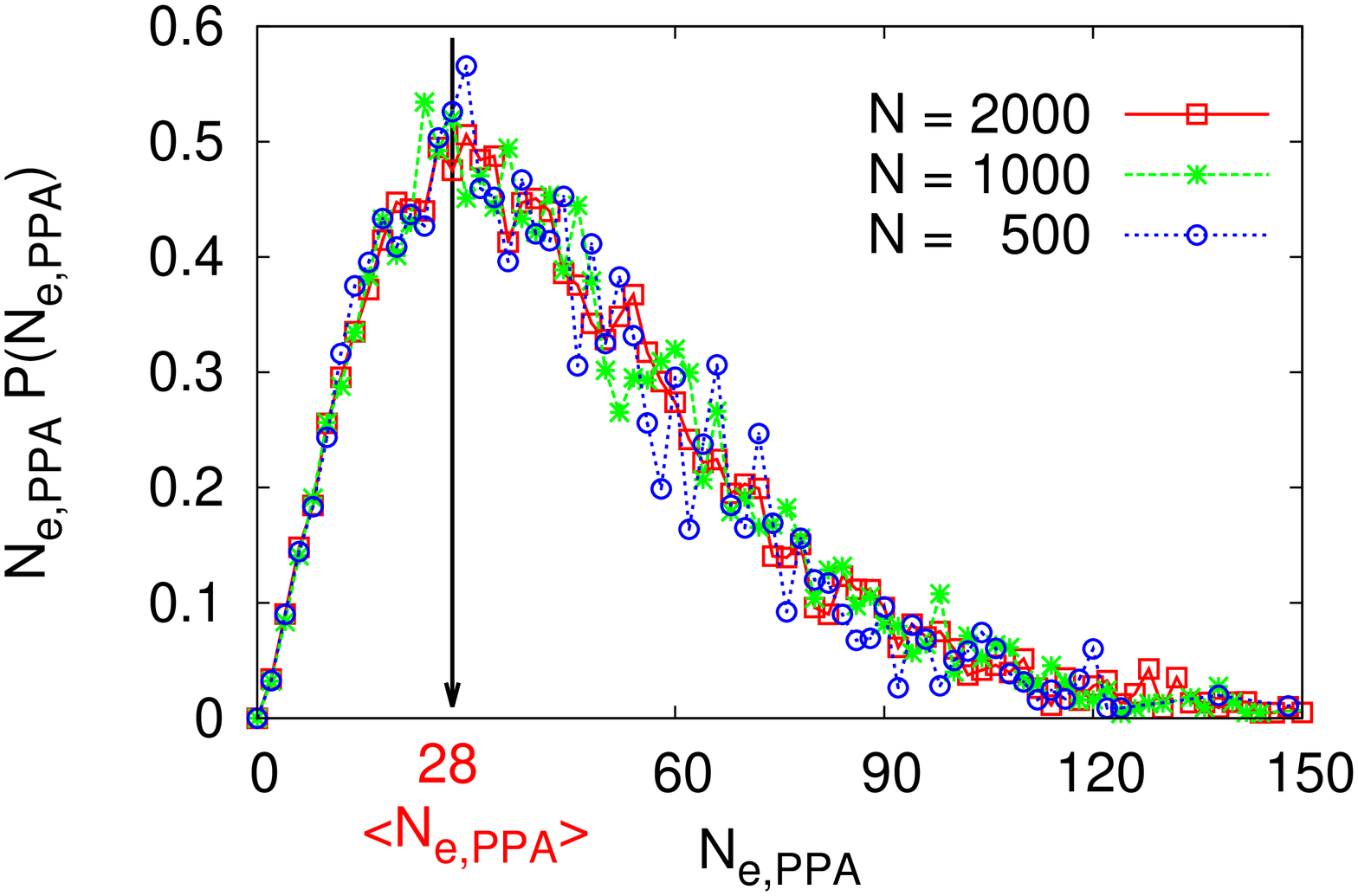} 
\caption{(a) Normalized probability distributions of bond length $\ell_b^{(pp)}$
of the primitive paths,
$P(\ell_b^{(pp)})$. (b) Normalized probability distributions of the entanglement
lengths $N_{e,PPA}$,  $P(N_{e,PPA})$. (c) Same data as in (b) 
but multiplied by $N_{e,PPA}$.
The peak of $N_{e,PPA}P(N_{e,PPA})$ indicates the estimate of entanglement length 
$\langle N_{e,PPA} \rangle \approx 28$.
Data are for $N=500$, $1000$, and $2000$, as indicated, and for $k_\theta=1.5$.}
\label{fig-ppa-dist}
\end{center}
\end{figure*}

  Results of the mean square internal distance $\langle R^2(s) \rangle$ for 
the original and the primitive paths of polymer chains in a melt with $k_\theta=1.5$ 
are shown in Fig.~\ref{fig-ppa-Rs2}. Since the endpoints of chains are fixed, one
should expect that results of $\langle R^2(s) \rangle$ for both paths approach to
the same value with increasing $s$. It is indeed seen in Fig.~\ref{fig-ppa-Rs2}.
If we use $\langle \cos \theta \rangle=\exp(-s\ell_{b}^{(pp)}/\ell_{p}^{(pp)})$ where
$\ell_p^{(pp)}/\ell_b=N_{e,PPA}/2$ with $N_{e,PPA} \approx 28$ 
in Eq.~(\ref{eq-Rs2-FRC}), we see that results of $\langle R^2(s) \rangle$ for 
the primitive path can still be well described by the FRC.
We also check the distributions of bond length $\ell_{b}^{(pp)}$ \{Eq.~(\ref{eq-bpp})\}
for the primitive paths
and show that the distribution is simply a normal (Gaussian) distribution 
of $x$ ($x=\ell_b^{(pp)}$) given by
\begin{eqnarray}
      P_N(x) = \frac{1}{\sqrt{2\pi \sigma^2(x)}}  
\exp\left(-\frac{(x-\langle x \rangle)^2}{2\sigma^2(x)} \right) \nonumber \\
{\rm and}  \qquad  \int_0^\infty dx P_N(x)=1 \,,
\label{eq-px}
\end{eqnarray}
where $\sigma^2(x)=\langle x^2 \rangle - \langle x \rangle^2$ is the standard deviation
of $x$, and $\langle x \rangle$ is the mean value of $x$ (Fig.~\ref{fig-ppa-dist}).
The distributions of the entanglement length $N_{e,PPA}$, $P_N(N_{e,PPA})$, for $N=2000$, $1000$, and $500$, and 
for $k_\theta=1.5$ are shown in Fig.~\ref{fig-ppa-dist}b. We see that 
$P_N(N_{e,PPA}=\ell_{K}^{(pp)}/\ell_{b}^{(pp)})$ does 
not depend on $N$. The position of the peak of 
$N_{e,PPA}P(N_{e,PPA})$ (fig.~\ref{fig-ppa-dist}c) 
corresponds to the 
estimate~\cite{Moreira2015} of $\langle N_{e,PPA} \rangle \approx 28$.
Results of $N_{e,PPA}$ through the PPA are listed in Table~\ref{table1}
for three different chain sizes and for $k_\theta=1.5$.

\section{Viscoelasticity}

The viscoelasticity 
of polymer melts is normally characterized by the stress relaxation
modulus $G(t)$ as a function of relaxation time $t$.  
For $t<\tau_e$, $G(t) \sim t^{-1/2}$ since the dynamics of
chains can be described by the Rouse model
while $G(t)$ reaches a plateau value $G_N^0=(4/5)(\rho k_BT/N_e)$ 
depending on the entanglement 
length, or the molecular weight between entanglements 
predicted by the reptation theory~\cite{Doi1986, Doi1980} for $\tau_e<t \ll \tau_d$ where 
chains are assumed to move in a tube-like regime due to entanglements.
Finally, entangled chains are relaxed for $t>\tau_d$ and 
$G(t)$ starts to deviate from the plateau.

In order to clarify whether the entanglement length $N_e$ estimated
from stresses $\sigma(t)$ using the standard expression of the plateau modulus
$G_N^0=(4/5)(\rho k_BT/N_e)$ is equivalent to $N_{e,PPA}$ determined 
through PPA mentioned in Sec.~\ref{PPA}, we perform MD simulations
to estimate the stress relaxation modulus $G(t)$.
Two methods are considered here. One is from the stress autocorrelation
function (SAF) of off-diagonal elements of the preaveraged stress tensor for 
fully equilibrated polymer melts~\cite{Lee2009, Lee2010}. 
The components of stress tensor taking the pairwise potential $U_{ij}$ and
the three-body potential $U_{ijk}$ into account are defined via the virial theorem:
\begin{widetext}
\begin{eqnarray}
  \sigma_{\alpha \beta} (t)=& -&\frac{1}{V} \left(
\sum_{i=1}^{n_cN} m_i v_i^{(\alpha)}v_i^{(\beta)} + \frac{1}{2} 
\sum_{i,j=1}^{n_cN}  f_{ij}^{(\alpha)} r_{ij}^{(\beta)} \right) \nonumber \\
&+&\frac{1}{6V}\sum_{i,j,k=1}^{n_cN}  
\left(\frac{r_{ij}^{(\alpha)} r_{ij}^{(\beta)}}{r_{ij}} \frac{\partial U_{ijk}}{\partial r_{ij}}
+\frac{r_{jk}^{(\alpha)} r_{jk}^{(\beta)}}{r_{jk}} \frac{\partial U_{ijk}}{\partial r_{jk}}+
\frac{r_{ki}^{(\alpha)} r_{ki}^{(\beta)}}{r_{ki}} \frac{\partial U_{ijk}}{\partial r_{ki}} \right)
\end{eqnarray}
\end{widetext}
where $m_i$ and $v_i^{\alpha}$ are the mass and the $\alpha$th component of
the velocity vector of the $i$th bead, respectively, and 
$f_{ij}^{(\alpha)} \equiv - r^{(\alpha)}_{ij}\frac{\partial U_{ij}}{\partial r_{ij}}$
is the $\alpha$th component of the force vector acting on the $i$th bead by
the $j$th bead.
Using the Green-Kubo relationship~\cite{Sen2005}, the stress
relaxation modulus 
\begin{equation}
     G(t)=(G_{xy}(t)+G_{xz}(t)+G_{yz}(t))/3
\label{eq-Gt-1}
\end{equation}
where the off-diagonal element $G_{\alpha \beta}(t)=(V/k_BT){\rm SAF}_{\alpha \beta}(t)$.
In order to reduce the strong noise in SAF~\cite{Lee2009,Lee2010}, 
${\rm SAF}_{\alpha \beta}(t)$ is defined by
\begin{equation}
     SAF_{\alpha \beta}(t)=\langle \bar{\sigma}_{\alpha \beta}(t)
  \bar{\sigma}_{\alpha \beta}(0) \rangle \,
\label{eq-SAF}
\end{equation}
where the preaveraged stress tensor
\begin{equation}
   \bar{\sigma}_{\alpha \beta}(t)=\frac{1}{N_t} \sum_{k=1}^{N_t}
  \sigma_{\alpha \beta}(t+k\delta t) \,.
\label{eq-stress}
\end{equation}
In our simulations, we choose $N_t=100$ MD steps with the time step
$\delta t=0.01[\tau]$. 

\begin{table}
\caption{Estimates of the entanglement lengths $N_e$ from relaxation plateau modulus
$G_N^0$ and $N_{e,PPA}$ from the primitive path analysis for polymer chains of 
sizes $N=2000$, $1000$, and $500$ in a melt and for $k_\theta=1.5$. }
\begin{tabular}{ccc}
\hline
 $N$  &  $N_e$(plateau) & $N_{e,PPA}$\\
\hline
\hline
$2000$  & $28 \pm 2$ &  $28.01\pm 1.06$\\
$1000$  & $26 \pm 3$ &  $28.30\pm 1.38$\\
$500$   &  $28 \pm 3$ &   $27.60\pm 1.45$\\
\hline
\end{tabular}
\label{table1}
\end{table}

The other method is to measure the normal stress decay $\sigma_{norm}(t)$
after deforming polymer chains in a melt by a small step strain elongation,
since linear viscoelastic properties are associated with near equilibrium
measurements of the system where the configurations of polymer chains are 
not moved far away from their equilibrium states. 
In our simulations, this is done 
by applying $N_{cycles}$ cycles of uniaxial elongation to deform the
simulation box with a strain rate $\tau_R^{-1}<\dot{\varepsilon}<\tau_e^{-1}$
(holding each chain in a tube-like regime) at each cycle such that at the end the simulation box is elongated in
the $x$-direction ($L_x=\lambda L$), but shrunk in the $y$-, $z$-directions
($L_y=L_z=L/\sqrt{\lambda}$). 
Here the volume of the simulation box is kept fixed, $V=L^3$, and
the stretch ratio $\lambda=(1.02)^{N_{cycles}}\approx 1.2$ with $N_{cycles}=9$
such that the system is in the linear viscoelastic regime. 
Using the stress-strain formulas for classical rubber elasticity~\cite{Treloar1986},
the stress relaxation modulus 
\begin{eqnarray}
   G(t)=\frac{\sigma_{norm}(t)}{\lambda^2-1/\lambda}
       =\frac{(\sigma_{xx}-\frac{1}{2}(\sigma_{yy}+\sigma_{zz})}
         {\lambda^2-1/\lambda} \,.
\label{eq-Gt-2}
\end{eqnarray}

\begin{figure}[t]
\begin{center}
(a)\includegraphics[width=0.30\textwidth,angle=270]{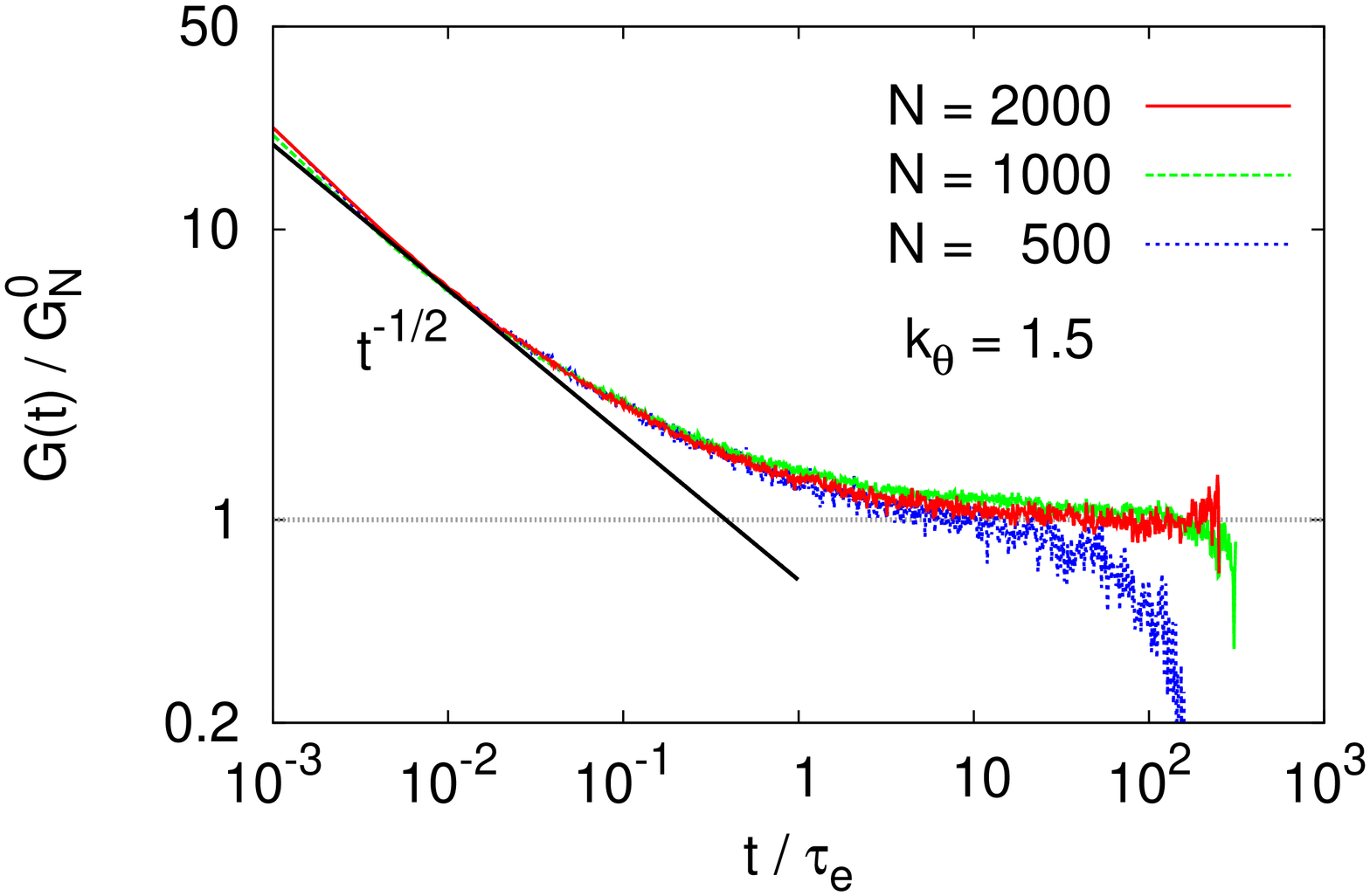} \hspace{0.4cm}
(b)\includegraphics[width=0.30\textwidth,angle=270]{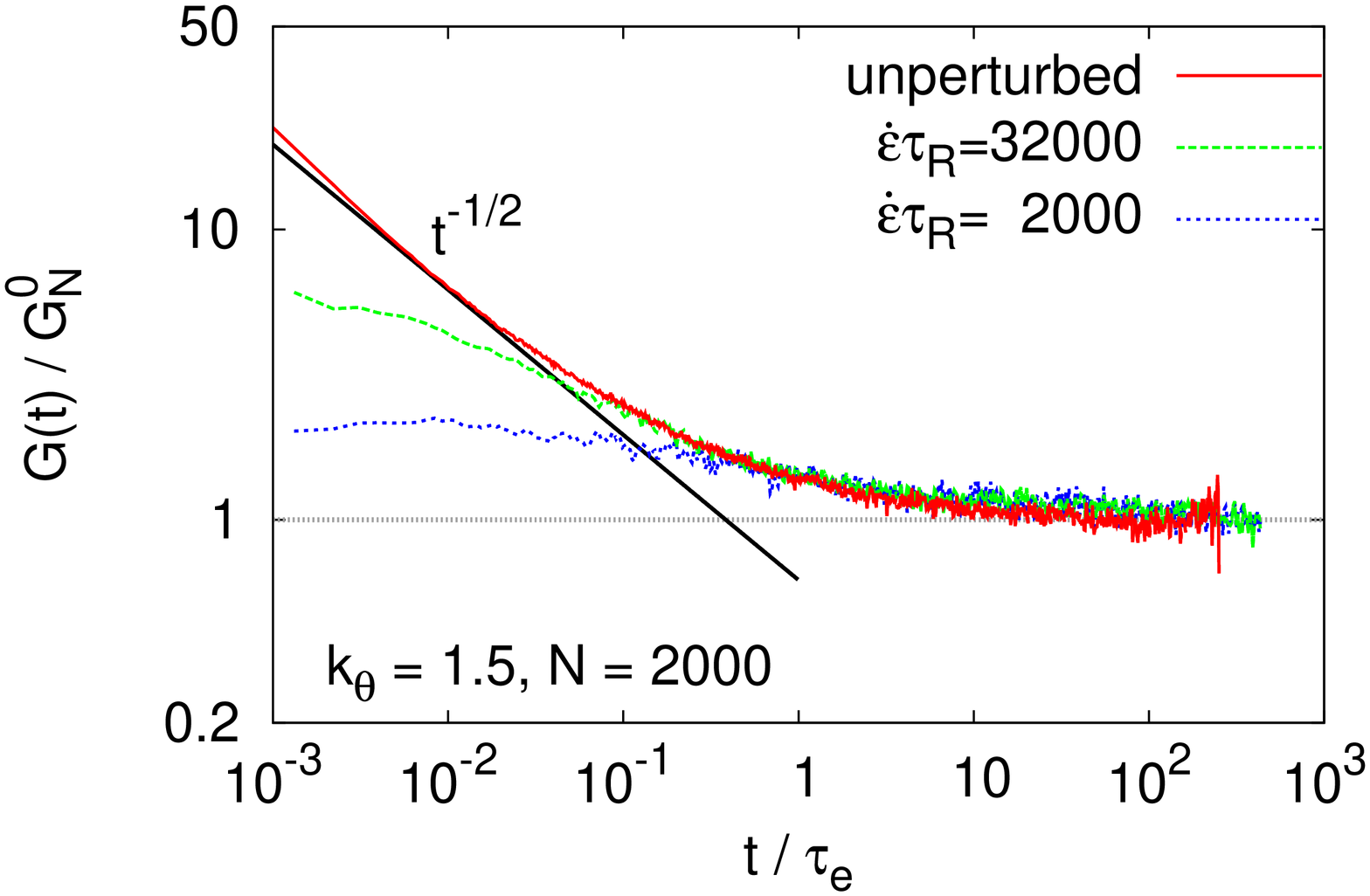}\\

\caption{(a) Stress relaxation modulus $G(t)$ scaled by $G_N^0=(4/5)\rho k_BT/N_e$
plotted as a function of $t$ from SAF \{Eq.~(\ref{eq-SAF})\} 
using the Green-Kubo relation.
(b) Same data for $N=2000$ as shown in (a), and $G(t)$ obtained 
from stress response to strain \{Eq.~(\ref{eq-Gt-2})\} after an
uniaxial elongation with two different strain rates $\dot{\varepsilon}$,
as indicated. Values of $N_e$ extracted from $G_N^0$ in (a) for three choices of $N$
are listed in Table~\ref{table1}.}
\label{fig-Gt}
\end{center}
\end{figure}

   Results of $G(t)$ scaled by $G_N^0$ with $N_e$ estimated 
by the plateau value of $G(t)$ are shown in Fig.~\ref{fig-Gt}. The estimates of 
$N_e$ are also listed in Table~\ref{table1}. They are in perfect agreement 
with the estimates through PPA within error bars.
In Fig.~\ref{fig-Gt}a, $G(t)$ is estimated from Eqs.~(\ref{eq-Gt-1})-(\ref{eq-stress})
for polymer melts consisting of $n_c=1000$ chains of sizes $N=500$, $1000$, and $2000$,
and for $k_\theta=1.5$. Due to the difference between microscopic structures of
independent equilibrated polymer melts, we observe that the plots of $G(t)$ as 
a function of $t$ show slightly different scenarios for different sets of data (not shown).
Therefore, besides taking the preaverage of $\sigma_{\alpha \beta}$ for the estimate
of $G(t)$, we shall also take the average of $G(t)$ over ${\cal O}(10)$ independent sets of data
although our systems are quite large.
For $t<\tau_e$, the scaling law $G(t) \sim t^{-1/2}$ predicted by the Rouse model is 
verified. As $t$ increases, the curves of $G(t)$ for three different sizes $N$ 
first reach a plateau for $\tau_e < t \ll \tau_d$, 
then start to deviate from it depending on the chain size $N$. Since
$\tau_d \sim N^{3.4}$, 
the range over which $G(t) \approx constant$ extends with
increasing $N$. However, in Fig.~\ref{fig-Gt}b, we only focus on the case of $N=2000$ and 
compare the results of $G(t)$ obtained from two different measurements,
Eqs~(\ref{eq-Gt-1}) and (\ref{eq-Gt-2}). For the second measurement, two values of the 
strain rate $\dot{\varepsilon}$ are chosen, $\dot{\varepsilon}\tau_R=2000$ and
$\dot{\varepsilon}\tau_R=32000$. We see that $G(t)$ only depends on $\dot{\varepsilon}$
for $t<t_e$. For $t>t_e$, results of $G(t)$ estimated from the normal stress tensor
$\sigma_{norm}(t)$  are consistent with the estimates from ${\rm SAF}(t)$.

\section{Conclusion}
    In this paper, we have studied bead-spring chains in a melt
at a monomer density $\rho=0.85$ by extensive molecular dynamics
simulations using the ESPResSo++ package~\cite{Espresso}.
We investigate the 
static and dynamic properties of polymer chains in a melt.
For fully equilibrated large polymer melts, we observe that
for moderately stiff chains ($k_\theta=1.5$), the ratio 
$\langle R_e^2 \rangle/\langle R_g^2 \rangle \approx 6$ as 
expected for ideal chains. 
For fully flexible chains ($k_\theta=0$), results of 
the mean square internal distance $\langle R^2(s) \rangle$ 
show remarkable deviations from the freely rotating chain model
describing the behavior of ideal chains,
while the deviations are diminished as the stiffness of chains
increases. For $k_\theta=1.5$, $\langle R^2(s)\rangle$ is in
perfect agreement with FRC up to $N \approx 800$, while 
a slight deviation occurs for $N>800$ due to the correlation hole
effect. Results of the probability distributions of reduced end-to-end 
distance $r_e =(R^2_e/\langle R_e^2 \rangle)^{1/2}$, and 
reduced gyration radius $r_g = (R^2_g/\langle R_g^2 \rangle)^{1/2}$
for polymer chains in a melt for various values of $N$ and for $k_\theta=1.5$
show the nice data collapse, and are described by
universal functions, Eqs.~(\ref{eq-Ge}) and (\ref{eq-Gg}), for ideal chains.
A detailed investigation of the standard structure factor $S_c(q)$
for single chains in a melt for $k_\theta=0$, $0.75$, and $1.5$ 
is also given. Results of $S_c(q)$ presented in a Kratky-plot show
that there exists a significant deviation from the Debye function
for Gaussian chains at the intermediate values of $q$ as 
observed by Wittmer et al.~\cite{Wittmer2007}, and Hsu~\cite{Hsu2014m},
while for $k_\theta=1.5$, it is very well described by the Debye function.
We have also seen that the probability distributions of bond angles
for $k_\theta>0$ have a bimodal form which is very different from the
distribution for ideal chains \{Eq.~(\ref{eq-ptheta})\}, 
and the positions of two peaks depend on the stiffness of chains.
All these findings support the idea that polymer chains in a melt are described
by ideal chains to some extent. The stiffer the chains of fixed 
size in a melt, the more ideal the chain.

   From our extensive molecular dynamics simulations, we have provided
evidence for the crossover behavior of the mean square
displacements $g_1(t)$, $g_2(t)$, and $g_3(t)$ between several
characteristic time scales, $\tau_0$, $\tau_e$, $\tau_R$, and $\tau_d$, as
predicted by the Rouse model, 
and the reptation theory~\cite{deGennes1979,Doi1986,Rubinstein2003}.
Especially, our results for $N=2000$ strongly support the reptation 
theory and the scaling
law $g_1(t)=g_2(t) \propto t^{1/4}$ for $\tau_e<t<\tau_R$.
We also see that the corresponding values of the mean square displacement
of a monomer, $g_1(t)$, at $\tau_0$, $\tau_e$, $\tau_R$, and $\tau_d$,
show that the theoretical predictions are not only verified qualitatively,
but also quantitatively.
The entanglement length $N_e$ determined from the 
estimates of $\tau_0$, $\tau_e$ and $\tau_d$ is not affected
by the chain size, and the estimates of $N_e$ are consistent with 
the estimates through the primitive path analysis, and the plateau modulus 
obtained from the stresses (Table~\ref{table1}).

    Also a direct comparison between the original paths and the primitive
paths (obtained through the primitive path analysis) of polymer chains 
in a melt is presented in this work. Results of the bond-bond
orientational correlation function and the mean square internal distance
of chains verify the assumption that chains behave as random walks 
along their primitive paths.
The Kuhn length $\ell_K^{(pp)}$ of the primitive
path is larger than $\ell_K$ of the original path.
The probability distribution of the average bond length
along the primitive paths shows a normal Gaussian distribution.
The peak of the first moment of the probability distribution of 
the entanglement length, $N_{e,PPA}P(N_{e,PPA})$ corresponds to the 
expected value of $\langle N_{e,PPA} \rangle$.

   Finally, the stress relaxation modulus $G(t)$ which describes
viscoelasticity of polymer melts is estimated.
From the stress autocorrelation function (SAF) of off-diagonal 
elements using the Green-Kubo relation, and the normal stress tensor 
after applying an uniaxial elongation but still keeping the system in 
a linear regime, we verify the Rouse behavior, $G(t) \sim t^{-1/2}$ for 
$t<\tau_e$. We also see that $G(t)$ reaches a plateau value, and
the plateau stays longer as the size of chains increases as predicted 
by the reptation theory~\cite{deGennes1979,Doi1986,Rubinstein2003}.
Moreover, we show that using the standard expression of plateau modulus 
$G_N^0=(4/5)(\rho k_BT/N_e)$, the estimate of the 
entanglement $N_e$ from the stresses is equivalent to $N_{e,PPA}$
through PPA.

 All our results show that the coarse-grained bead-spring model
is an ideal model for understanding the properties of fully
equilibrated polymer chains in a melt from various aspects. It marks a
good compromise between chain flexibility, and small entanglement length.
While the flexibility allows for relatively large timesteps and the application
of recently developed equilibration schemes the moderate stiffness warrants small
deviations from ideality and at the same time relatively small entanglement lengths,
which are decisive for comparably small, though still huge, relaxation times.
Therefore, we expect that this model can serve as an optimal test case, where one can gain
insight into non-linear viscoelasticity regime for large
polymer melts by non-equilibrium molecular dynamics simulations.\\

\noindent
{\bf ACKNOWLEDGMENT}\\

  This work has been supported by European Research Council under the European 
Union's Seventh Framework Programme (FP7/2007-2013)/ERC Grant Agreement 
No.~340906-MOLPROCOMP. We are grateful to G.~S.~Grest for stimulating 
discussions. We also thank G.~Zhang for providing the configurations of 
large equilibrated polymer melt in equilibrium, K. Ch. Daoulas for 
helpful discussion, T. Stuehn and L. A. Moreira for assistance with the ESPResSo++
package, and A. C. Fogarty for a critical reading of the manuscript.
We are also grateful to the NIC J\"ulich for a generous
grant of computing time at the J\"ulich Supercomputing
Centre (JSC), and the Max Planck Computing and Data Facility (MPCDF).\\

\noindent
{\bf APPENDIX}\\

 Polymer melts consisting of $n_c$ chains of $N$ monomers
in a melt are described by the standard bead-spring
model~\cite{Kremer1990} at a volume fraction $\phi=0.85$ for our work here.
In this model, the excluded volume interactions between bonded and non-bonded
monomers at a distance $r$ are considered by the truncated and shifted Lennard-Jones (LJ) potential,
i.e., Weeks-Chandler-Andersen (WCA) potential,
\begin{eqnarray}
U_{\rm LJ}(r)=\left\{\begin{array}{ll}
 4\varepsilon\left[\left(\frac{\sigma}{r}\right)^{12}-\left(\frac{\sigma}{r}\right)^{6}
+\frac{1}{4} \right] & \;, \, r \le r_{\rm cut} \\
0 &\;, \, r>r_{\rm cut} 
\end{array} \right. \;
\end{eqnarray}
where $\varepsilon$ denotes the pairwise interaction energy, and $r_{\rm cut}=2^{1/6}\sigma$ is
a cut-off such that $U_{\rm LJ}(r_{\rm cut})=0=min. \{U_{\rm LJ}(r)\}$. The bond length
$\mid \vec{b}_j \mid=\mid \vec{r}_{j+1}-\vec{r}_j \mid$
between any two connected monomers $j$ and $j+1$ of size $\sigma$ and mass
$m$ along a chain
is controlled by the
finitely extensible nonlinear elastic (FENE) potential
\begin{eqnarray}
  U_{\rm FENE}(r) = \left\{\begin{array}{ll}
 -\frac{k}{2}R_0^2 \ln \left[1-\left(\frac{r}{R_0}\right)^2 \right] & \;,\, r \le R_0 \\
 \infty & \;, \, r>R_0 
\end{array} \right . \;
\end{eqnarray}
where the force constant $k=30\varepsilon/\sigma^2$, and the maximum value of
bond length $R_0=1.5\sigma$.
The chain stiffness is dominated by the bending potential depending
on the bond angle $\theta$ between the sequential
bonds along a chain itself and the strength of the bending factor $k_\theta$,
\begin{equation}
  U_{\rm bend}(\theta)=k_\theta(1-\cos \theta) \
\end{equation}

The ESPResSo++ package~\cite{Espresso} is used to perform the 
standard MD with Langevin thermostat
including a friction constant $\Gamma=0.5 \tau^{-1}$
where $\tau=\sigma(m/\varepsilon)^{1/2}$, and random force $\vec{f}_i^R$ that
\begin{equation}
  m \frac{d^2 \vec{r}_i}{dt^2} = -\bigtriangledown(U_{\rm LJ}+U_{\rm FENE}
+ U_{\rm bend}) -\Gamma \frac{d \vec{r}_i}{dt}+\vec{f}_i^R(t)
\end{equation}
and
\begin{equation}
 \langle \vec{f}_i^R(t) \cdot \vec{f}_j^R(t') \rangle = 
6 k_BT \Gamma \delta_{ij} \delta(t-t') \, .
\end{equation}
Here the temperature $T=1 \varepsilon/k_B$, $K_B$ is the Boltzmann factor, and
$\sigma=m=1$, and the basic time step $\Delta t$ for the integration 
is $0.01 \tau$ throughout the whole paper.
In the primitive path analysis \{Sec.~\ref{PPA}\}, 
we set the temperature $T=0.001 \varepsilon/k_B$ (close to zero), the basic 
time step $\Delta t=0.006$, the friction constant $\Gamma=20 \tau^{-1}$ during the 
first $10^3$  MD steps, and $\Gamma=0.5 \tau^{-1}$ after the 
first $10^{3}$ steps~\cite{Sukumaran2005,Moreira2015}.


\begin{thebibliography}{53}
\expandafter\ifx\csname natexlab\endcsname\relax\def\natexlab#1{#1}\fi
\expandafter\ifx\csname bibnamefont\endcsname\relax
  \def\bibnamefont#1{#1}\fi
\expandafter\ifx\csname bibfnamefont\endcsname\relax
  \def\bibfnamefont#1{#1}\fi
\expandafter\ifx\csname citenamefont\endcsname\relax
  \def\citenamefont#1{#1}\fi
\expandafter\ifx\csname url\endcsname\relax
  \def\url#1{\texttt{#1}}\fi
\expandafter\ifx\csname urlprefix\endcsname\relax\def\urlprefix{URL }\fi
\providecommand{\bibinfo}[2]{#2}
\providecommand{\eprint}[2][]{\url{#2}}

\bibitem[{\citenamefont{de~Gennes}(1979)}]{deGennes1979}
\bibinfo{author}{\bibfnamefont{P.~G.} \bibnamefont{de~Gennes}},
  \emph{\bibinfo{title}{Scaling Concepts in polymer physics}}
  (\bibinfo{publisher}{Cornell University Press: Itharca, New York},
  \bibinfo{year}{1979}).

\bibitem[{\citenamefont{Yamakawa}(1971)}]{Yamakawa1971}
\bibinfo{author}{\bibfnamefont{H.}~\bibnamefont{Yamakawa}},
  \emph{\bibinfo{title}{Modern theory of polymer solutions}}
  (\bibinfo{publisher}{Harper and Row, New York}, \bibinfo{year}{1971}).

\bibitem[{\citenamefont{Wittmer et~al.}(2007)\citenamefont{Wittmer, Beckrich,
  Johner, Semenov, Obukhov, Mayer, and Baschnagel}}]{Wittmer2007}
\bibinfo{author}{\bibfnamefont{J.~P.} \bibnamefont{Wittmer}},
  \bibinfo{author}{\bibfnamefont{P.}~\bibnamefont{Beckrich}},
  \bibinfo{author}{\bibfnamefont{A.}~\bibnamefont{Johner}},
  \bibinfo{author}{\bibfnamefont{A.~N.} \bibnamefont{Semenov}},
  \bibinfo{author}{\bibfnamefont{S.~P.} \bibnamefont{Obukhov}},
  \bibinfo{author}{\bibfnamefont{H.}~\bibnamefont{Mayer}}, \bibnamefont{and}
  \bibinfo{author}{\bibfnamefont{J.}~\bibnamefont{Baschnagel}},
  \bibinfo{journal}{EPL} \textbf{\bibinfo{volume}{77}}, \bibinfo{pages}{56003}
  (\bibinfo{year}{2007}).

\bibitem[{\citenamefont{Wittmer et~al.}(2011)\citenamefont{Wittmer, Cavallo,
  H.~Xu, Poli\'{n}ska, Schulmann, Meyer, Farago, Johner, Obukhov, and
  Baschnagel}}]{Wittmer2011}
\bibinfo{author}{\bibfnamefont{J.~P.} \bibnamefont{Wittmer}},
  \bibinfo{author}{\bibfnamefont{A.}~\bibnamefont{Cavallo}},
  \bibinfo{author}{\bibfnamefont{J.~E.~Z.} \bibnamefont{H.~Xu}},
  \bibinfo{author}{\bibfnamefont{P.}~\bibnamefont{Poli\'{n}ska}},
  \bibinfo{author}{\bibfnamefont{N.}~\bibnamefont{Schulmann}},
  \bibinfo{author}{\bibfnamefont{H.}~\bibnamefont{Meyer}},
  \bibinfo{author}{\bibfnamefont{J.}~\bibnamefont{Farago}},
  \bibinfo{author}{\bibfnamefont{A.}~\bibnamefont{Johner}},
  \bibinfo{author}{\bibfnamefont{S.~P.} \bibnamefont{Obukhov}},
  \bibnamefont{and}
  \bibinfo{author}{\bibfnamefont{J.}~\bibnamefont{Baschnagel}},
  \bibinfo{journal}{J. Stat. Phys.} \textbf{\bibinfo{volume}{145}},
  \bibinfo{pages}{1017} (\bibinfo{year}{2011}).

\bibitem[{\citenamefont{Hsu}(2014{\natexlab{a}})}]{Hsu2014}
\bibinfo{author}{\bibfnamefont{H.-P.} \bibnamefont{Hsu}}, \bibinfo{journal}{J.
  Chem. Phys.} \textbf{\bibinfo{volume}{141}}, \bibinfo{pages}{164903}
  (\bibinfo{year}{2014}{\natexlab{a}}).

\bibitem[{\citenamefont{Doi and Edwards}(1986)}]{Doi1986}
\bibinfo{author}{\bibfnamefont{M.}~\bibnamefont{Doi}} \bibnamefont{and}
  \bibinfo{author}{\bibfnamefont{S.}~\bibnamefont{Edwards}},
  \emph{\bibinfo{title}{The theory of polymer dynamics}}
  (\bibinfo{publisher}{Oxford University Press: New York},
  \bibinfo{year}{1986}).

\bibitem[{\citenamefont{Kalathi et~al.}(2014)\citenamefont{Kalathi, Kumar,
  Rubinstein, and Grest}}]{Kalathi2014}
\bibinfo{author}{\bibfnamefont{J.~T.} \bibnamefont{Kalathi}},
  \bibinfo{author}{\bibfnamefont{S.~K.} \bibnamefont{Kumar}},
  \bibinfo{author}{\bibfnamefont{M.}~\bibnamefont{Rubinstein}},
  \bibnamefont{and} \bibinfo{author}{\bibfnamefont{G.~S.} \bibnamefont{Grest}},
  \bibinfo{journal}{Macromolecules} \textbf{\bibinfo{volume}{47}},
  \bibinfo{pages}{6925} (\bibinfo{year}{2014}).

\bibitem[{\citenamefont{Kalathi et~al.}(2015)\citenamefont{Kalathi, Kumar,
  Rubinstein, and Grest}}]{Kalathi2015}
\bibinfo{author}{\bibfnamefont{J.~T.} \bibnamefont{Kalathi}},
  \bibinfo{author}{\bibfnamefont{S.~K.} \bibnamefont{Kumar}},
  \bibinfo{author}{\bibfnamefont{M.}~\bibnamefont{Rubinstein}},
  \bibnamefont{and} \bibinfo{author}{\bibfnamefont{G.~S.} \bibnamefont{Grest}},
  \bibinfo{journal}{Soft Matter} \textbf{\bibinfo{volume}{11}},
  \bibinfo{pages}{4123} (\bibinfo{year}{2015}).

\bibitem[{\citenamefont{Rubinstein and Colby}(2003)}]{Rubinstein2003}
\bibinfo{author}{\bibfnamefont{M.}~\bibnamefont{Rubinstein}} \bibnamefont{and}
  \bibinfo{author}{\bibfnamefont{R.~H.} \bibnamefont{Colby}},
  \emph{\bibinfo{title}{Polymer Physics}} (\bibinfo{publisher}{Oxford
  University Press, Oxford}, \bibinfo{year}{2003}).

\bibitem[{\citenamefont{Baumg\"artner and Binder}(1981)}]{Baumgaertner1981}
\bibinfo{author}{\bibfnamefont{A.}~\bibnamefont{Baumg\"artner}}
  \bibnamefont{and} \bibinfo{author}{\bibfnamefont{K.}~\bibnamefont{Binder}},
  \bibinfo{journal}{J. Chem. Phys.} \textbf{\bibinfo{volume}{75}},
  \bibinfo{pages}{2994} (\bibinfo{year}{1981}).

\bibitem[{\citenamefont{Kremer and Grest}(1990)}]{Kremer1990}
\bibinfo{author}{\bibfnamefont{K.}~\bibnamefont{Kremer}} \bibnamefont{and}
  \bibinfo{author}{\bibfnamefont{G.~S.} \bibnamefont{Grest}},
  \bibinfo{journal}{J. Chem. Phys.} \textbf{\bibinfo{volume}{92}},
  \bibinfo{pages}{5057} (\bibinfo{year}{1990}).

\bibitem[{\citenamefont{Paul et~al.}(1991)\citenamefont{Paul, Binder, Heermann,
  and Kremer}}]{Paul1991}
\bibinfo{author}{\bibfnamefont{W.}~\bibnamefont{Paul}},
  \bibinfo{author}{\bibfnamefont{K.}~\bibnamefont{Binder}},
  \bibinfo{author}{\bibfnamefont{D.~W.} \bibnamefont{Heermann}},
  \bibnamefont{and} \bibinfo{author}{\bibfnamefont{K.}~\bibnamefont{Kremer}},
  \bibinfo{journal}{J. Chem. Phys.} \textbf{\bibinfo{volume}{95}},
  \bibinfo{pages}{7726} (\bibinfo{year}{1991}).

\bibitem[{\citenamefont{Wittmer et~al.}(1992)\citenamefont{Wittmer, Paul, and
  Binder}}]{Wittmer1992}
\bibinfo{author}{\bibfnamefont{J.}~\bibnamefont{Wittmer}},
  \bibinfo{author}{\bibfnamefont{W.}~\bibnamefont{Paul}}, \bibnamefont{and}
  \bibinfo{author}{\bibfnamefont{K.}~\bibnamefont{Binder}},
  \bibinfo{journal}{Macromolecules} \textbf{\bibinfo{volume}{25}},
  \bibinfo{pages}{7211} (\bibinfo{year}{1992}).

\bibitem[{\citenamefont{k.~Kremer and Grest}(1992)}]{Kremer1992}
\bibinfo{author}{\bibnamefont{k.~Kremer}} \bibnamefont{and}
  \bibinfo{author}{\bibfnamefont{G.~S.} \bibnamefont{Grest}},
  \bibinfo{journal}{J. Chem. Soc. Faraday Trans} \textbf{\bibinfo{volume}{88}},
  \bibinfo{pages}{1707} (\bibinfo{year}{1992}).

\bibitem[{\citenamefont{Kopf et~al.}(1997)\citenamefont{Kopf, D\"unweg, and
  Paul}}]{Kopf1997}
\bibinfo{author}{\bibfnamefont{A.}~\bibnamefont{Kopf}},
  \bibinfo{author}{\bibfnamefont{B.}~\bibnamefont{D\"unweg}}, \bibnamefont{and}
  \bibinfo{author}{\bibfnamefont{W.}~\bibnamefont{Paul}}, \bibinfo{journal}{J.
  Chem. Phys.} \textbf{\bibinfo{volume}{107}}, \bibinfo{pages}{6945}
  (\bibinfo{year}{1997}).

\bibitem[{\citenamefont{Zhang et~al.}(2014)\citenamefont{Zhang, Moreira,
  Stuehn, Daoulas, and Kremer}}]{Zhang2014}
\bibinfo{author}{\bibfnamefont{G.}~\bibnamefont{Zhang}},
  \bibinfo{author}{\bibfnamefont{L.~A.} \bibnamefont{Moreira}},
  \bibinfo{author}{\bibfnamefont{T.}~\bibnamefont{Stuehn}},
  \bibinfo{author}{\bibfnamefont{K.~C.} \bibnamefont{Daoulas}},
  \bibnamefont{and} \bibinfo{author}{\bibfnamefont{K.}~\bibnamefont{Kremer}},
  \bibinfo{journal}{ACS Macro Lett.} \textbf{\bibinfo{volume}{3}},
  \bibinfo{pages}{198} (\bibinfo{year}{2014}).

\bibitem[{\citenamefont{Vettorel et~al.}(2010)\citenamefont{Vettorel, Besold,
  and Kremer}}]{Vettorel2010}
\bibinfo{author}{\bibfnamefont{T.}~\bibnamefont{Vettorel}},
  \bibinfo{author}{\bibfnamefont{G.}~\bibnamefont{Besold}}, \bibnamefont{and}
  \bibinfo{author}{\bibfnamefont{K.}~\bibnamefont{Kremer}},
  \bibinfo{journal}{Soft Matter} \textbf{\bibinfo{volume}{6}},
  \bibinfo{pages}{2282} (\bibinfo{year}{2010}).

\bibitem[{\citenamefont{Zhang et~al.}(2013)\citenamefont{Zhang, Daoulas, and
  Kremer}}]{Zhang2013}
\bibinfo{author}{\bibfnamefont{G.}~\bibnamefont{Zhang}},
  \bibinfo{author}{\bibfnamefont{K.~C.} \bibnamefont{Daoulas}},
  \bibnamefont{and} \bibinfo{author}{\bibfnamefont{K.}~\bibnamefont{Kremer}},
  \bibinfo{journal}{Macromol. Chem. Phys.} \textbf{\bibinfo{volume}{214}},
  \bibinfo{pages}{214} (\bibinfo{year}{2013}).

\bibitem[{\citenamefont{Everaers et~al.}(2004)\citenamefont{Everaers,
  Sukumaran, Grest, Svaneborg, Sivasubramanian, and Kremer}}]{Everaers2004}
\bibinfo{author}{\bibfnamefont{R.}~\bibnamefont{Everaers}},
  \bibinfo{author}{\bibfnamefont{S.~K.} \bibnamefont{Sukumaran}},
  \bibinfo{author}{\bibfnamefont{G.~S.} \bibnamefont{Grest}},
  \bibinfo{author}{\bibfnamefont{C.}~\bibnamefont{Svaneborg}},
  \bibinfo{author}{\bibfnamefont{A.}~\bibnamefont{Sivasubramanian}},
  \bibnamefont{and} \bibinfo{author}{\bibfnamefont{K.}~\bibnamefont{Kremer}},
  \bibinfo{journal}{Science} \textbf{\bibinfo{volume}{303}},
  \bibinfo{pages}{823} (\bibinfo{year}{2004}).

\bibitem[{\citenamefont{Fujita and Norisuye}(1970)}]{Fujita1970}
\bibinfo{author}{\bibfnamefont{H.}~\bibnamefont{Fujita}} \bibnamefont{and}
  \bibinfo{author}{\bibfnamefont{T.}~\bibnamefont{Norisuye}},
  \bibinfo{journal}{J. Chem. Phys.} \textbf{\bibinfo{volume}{52}},
  \bibinfo{pages}{1115} (\bibinfo{year}{1970}).

\bibitem[{\citenamefont{Denton and Schmidt}(2002)}]{Denton2002}
\bibinfo{author}{\bibfnamefont{A.~R.} \bibnamefont{Denton}} \bibnamefont{and}
  \bibinfo{author}{\bibfnamefont{M.}~\bibnamefont{Schmidt}},
  \bibinfo{journal}{J. Phys.: Condens. Matter} \textbf{\bibinfo{volume}{14}},
  \bibinfo{pages}{12051} (\bibinfo{year}{2002}).

\bibitem[{\citenamefont{Hsu}(2014{\natexlab{b}})}]{Hsu2014m}
\bibinfo{author}{\bibfnamefont{H.-P.} \bibnamefont{Hsu}}, \bibinfo{journal}{J.
  Chem. Phys.} \textbf{\bibinfo{volume}{141}}, \bibinfo{pages}{234901}
  (\bibinfo{year}{2014}{\natexlab{b}}).

\bibitem[{\citenamefont{Lhuillier}(1988)}]{Lhuillier1988}
\bibinfo{author}{\bibfnamefont{D.}~\bibnamefont{Lhuillier}},
  \bibinfo{journal}{J. Phys. France} \textbf{\bibinfo{volume}{49}},
  \bibinfo{pages}{705} (\bibinfo{year}{1988}).

\bibitem[{\citenamefont{des Cloizeaux}(1975)}]{Cloizeaux1975}
\bibinfo{author}{\bibfnamefont{J.}~\bibnamefont{des Cloizeaux}},
  \bibinfo{journal}{J. Phys. France} \textbf{\bibinfo{volume}{36}},
  \bibinfo{pages}{281} (\bibinfo{year}{1975}).

\bibitem[{\citenamefont{Fisher}(1966)}]{Fisher1966}
\bibinfo{author}{\bibfnamefont{M.~E.} \bibnamefont{Fisher}},
  \bibinfo{journal}{J. Chem. Phys.} \textbf{\bibinfo{volume}{44}},
  \bibinfo{pages}{616} (\bibinfo{year}{1966}).

\bibitem[{\citenamefont{Flory}(1969)}]{Flory1969}
\bibinfo{author}{\bibfnamefont{P.~J.} \bibnamefont{Flory}},
  \emph{\bibinfo{title}{Statistical Mechanics of Chain Molecules}}
  (\bibinfo{publisher}{Wiley, New York}, \bibinfo{year}{1969}).

\bibitem[{\citenamefont{Grosbeg and Khokhlov}(1994)}]{Grosberg1994}
\bibinfo{author}{\bibfnamefont{A.~Y.} \bibnamefont{Grosbeg}} \bibnamefont{and}
  \bibinfo{author}{\bibfnamefont{A.~R.} \bibnamefont{Khokhlov}},
  \emph{\bibinfo{title}{Statistical Physics of Macromolecules}}
  (\bibinfo{publisher}{AIP Press, NY}, \bibinfo{year}{1994}).

\bibitem[{\citenamefont{Auhl et~al.}(2003)\citenamefont{Auhl, Everaers, Grest,
  Kremer, and Plimpton}}]{Auhl2003}
\bibinfo{author}{\bibfnamefont{R.}~\bibnamefont{Auhl}},
  \bibinfo{author}{\bibfnamefont{R.}~\bibnamefont{Everaers}},
  \bibinfo{author}{\bibfnamefont{G.~S.} \bibnamefont{Grest}},
  \bibinfo{author}{\bibfnamefont{K.}~\bibnamefont{Kremer}}, \bibnamefont{and}
  \bibinfo{author}{\bibfnamefont{S.~J.} \bibnamefont{Plimpton}},
  \bibinfo{journal}{J. Chem. Phys.} \textbf{\bibinfo{volume}{119}},
  \bibinfo{pages}{12718} (\bibinfo{year}{2003}).

\bibitem[{\citenamefont{Hsu et~al.}(2010)\citenamefont{Hsu, Paul, and
  Binder}}]{Hsu2010}
\bibinfo{author}{\bibfnamefont{H.-P.} \bibnamefont{Hsu}},
  \bibinfo{author}{\bibfnamefont{W.}~\bibnamefont{Paul}}, \bibnamefont{and}
  \bibinfo{author}{\bibfnamefont{K.}~\bibnamefont{Binder}},
  \bibinfo{journal}{Macromolecules} \textbf{\bibinfo{volume}{20}},
  \bibinfo{pages}{510} (\bibinfo{year}{2010}).

\bibitem[{\citenamefont{Cloizeaux and Jannink}(1990)}]{Cloizeaux1990}
\bibinfo{author}{\bibfnamefont{J.~D.} \bibnamefont{Cloizeaux}}
  \bibnamefont{and} \bibinfo{author}{\bibfnamefont{G.}~\bibnamefont{Jannink}},
  \emph{\bibinfo{title}{Polymers in Solution: Their Modeling and Structure}}
  (\bibinfo{publisher}{Clarendon, Oxford}, \bibinfo{year}{1990}).

\bibitem[{\citenamefont{Sch{\"a}fer}(1999)}]{Schaefer1999}
\bibinfo{author}{\bibfnamefont{L.}~\bibnamefont{Sch{\"a}fer}},
  \emph{\bibinfo{title}{Excluded Volume Effects in Polymer Solutions as
  Explained by the Renormalization Group}} (\bibinfo{publisher}{Springer,
  Berlin}, \bibinfo{year}{1999}).

\bibitem[{\citenamefont{Higgins and Benoit}(1994)}]{Higgins1994}
\bibinfo{author}{\bibfnamefont{J.~S.} \bibnamefont{Higgins}} \bibnamefont{and}
  \bibinfo{author}{\bibfnamefont{H.~C.} \bibnamefont{Benoit}},
  \emph{\bibinfo{title}{Polymers and Neutron Scattering}}
  (\bibinfo{publisher}{Clarendon, Oxford}, \bibinfo{year}{1994}).

\bibitem[{\citenamefont{Doi}(1983)}]{Doi1983}
\bibinfo{author}{\bibfnamefont{M.}~\bibnamefont{Doi}}, \bibinfo{journal}{J.
  Polym. Sci. Polym. Phys. Ed.} \textbf{\bibinfo{volume}{21}},
  \bibinfo{pages}{667} (\bibinfo{year}{1983}).

\bibitem[{\citenamefont{Milner and McLeish}(1998)}]{Milner1998}
\bibinfo{author}{\bibfnamefont{S.~T.} \bibnamefont{Milner}} \bibnamefont{and}
  \bibinfo{author}{\bibfnamefont{T.~C.~B.} \bibnamefont{McLeish}},
  \bibinfo{journal}{Phys. Rev. Lett.} \textbf{\bibinfo{volume}{81}},
  \bibinfo{pages}{725} (\bibinfo{year}{1998}).

\bibitem[{\citenamefont{McLeish}(2002)}]{McLeish2002}
\bibinfo{author}{\bibfnamefont{T.~C.~B.} \bibnamefont{McLeish}},
  \bibinfo{journal}{Adv. Phys.} \textbf{\bibinfo{volume}{51}},
  \bibinfo{pages}{1379} (\bibinfo{year}{2002}).

\bibitem[{\citenamefont{Likhtman and McLeish}(2002)}]{Likhtman2002}
\bibinfo{author}{\bibfnamefont{A.~E.} \bibnamefont{Likhtman}} \bibnamefont{and}
  \bibinfo{author}{\bibfnamefont{T.~C.~B.} \bibnamefont{McLeish}},
  \bibinfo{journal}{Macromolecules} \textbf{\bibinfo{volume}{35}},
  \bibinfo{pages}{6332} (\bibinfo{year}{2002}).

\bibitem[{\citenamefont{P\"utz et~al.}(2000)\citenamefont{P\"utz, Kremer, and
  Grest}}]{Puetz2000}
\bibinfo{author}{\bibfnamefont{M.}~\bibnamefont{P\"utz}},
  \bibinfo{author}{\bibfnamefont{K.}~\bibnamefont{Kremer}}, \bibnamefont{and}
  \bibinfo{author}{\bibfnamefont{G.~S.} \bibnamefont{Grest}},
  \bibinfo{journal}{Europhys. Lett.} \textbf{\bibinfo{volume}{49}},
  \bibinfo{pages}{735} (\bibinfo{year}{2000}).

\bibitem[{\citenamefont{Anderson et~al.}(2008)\citenamefont{Anderson, Lorenz,
  and Travesset}}]{Anderson2008}
\bibinfo{author}{\bibfnamefont{J.~A.} \bibnamefont{Anderson}},
  \bibinfo{author}{\bibfnamefont{C.}~\bibnamefont{Lorenz}}, \bibnamefont{and}
  \bibinfo{author}{\bibfnamefont{A.}~\bibnamefont{Travesset}},
  \bibinfo{journal}{J. Comput. Phys.} \textbf{\bibinfo{volume}{227}},
  \bibinfo{pages}{5342} (\bibinfo{year}{2008}).

\bibitem[{\citenamefont{Glaser et~al.}(2015)\citenamefont{Glaser, Nguyen,
  Anderson, Lui, Spiga, adn D.~C.~Morse, and Glotzer}}]{Glaser2015}
\bibinfo{author}{\bibfnamefont{J.}~\bibnamefont{Glaser}},
  \bibinfo{author}{\bibfnamefont{T.~D.} \bibnamefont{Nguyen}},
  \bibinfo{author}{\bibfnamefont{J.~A.} \bibnamefont{Anderson}},
  \bibinfo{author}{\bibfnamefont{P.}~\bibnamefont{Lui}},
  \bibinfo{author}{\bibfnamefont{F.}~\bibnamefont{Spiga}},
  \bibinfo{author}{\bibfnamefont{J.~A.~M.} \bibnamefont{adn D.~C.~Morse}},
  \bibnamefont{and} \bibinfo{author}{\bibfnamefont{S.~C.}
  \bibnamefont{Glotzer}}, \bibinfo{journal}{Comput. Phys. Comm.}
  \textbf{\bibinfo{volume}{192}}, \bibinfo{pages}{97} (\bibinfo{year}{2015}).

\bibitem[{\citenamefont{Edwards}(1967)}]{Edwards1967}
\bibinfo{author}{\bibfnamefont{S.~F.} \bibnamefont{Edwards}},
  \bibinfo{journal}{Proc. Phys. Soc.} \textbf{\bibinfo{volume}{91}},
  \bibinfo{pages}{513} (\bibinfo{year}{1967}).

\bibitem[{\citenamefont{Sukumaran et~al.}(2005)\citenamefont{Sukumaran, Grest,
  Kremer, and Everaers}}]{Sukumaran2005}
\bibinfo{author}{\bibfnamefont{S.~K.} \bibnamefont{Sukumaran}},
  \bibinfo{author}{\bibfnamefont{G.~S.} \bibnamefont{Grest}},
  \bibinfo{author}{\bibfnamefont{K.}~\bibnamefont{Kremer}}, \bibnamefont{and}
  \bibinfo{author}{\bibfnamefont{R.}~\bibnamefont{Everaers}},
  \bibinfo{journal}{J. Polym. Sci. B} \textbf{\bibinfo{volume}{43}},
  \bibinfo{pages}{917} (\bibinfo{year}{2005}).

\bibitem[{\citenamefont{Kr\"oger}(2005)}]{Kroeger2005}
\bibinfo{author}{\bibfnamefont{M.}~\bibnamefont{Kr\"oger}},
  \bibinfo{journal}{Comput. Phys. Comm.} \textbf{\bibinfo{volume}{168}},
  \bibinfo{pages}{209} (\bibinfo{year}{2005}).

\bibitem[{\citenamefont{Shanbhag and Larson}(2005)}]{Shanbhag2005}
\bibinfo{author}{\bibfnamefont{S.}~\bibnamefont{Shanbhag}} \bibnamefont{and}
  \bibinfo{author}{\bibfnamefont{R.~G.} \bibnamefont{Larson}},
  \bibinfo{journal}{Phys. Rev. Lett.} \textbf{\bibinfo{volume}{94}},
  \bibinfo{pages}{076001} (\bibinfo{year}{2005}).

\bibitem[{\citenamefont{Tzoumanekas and Theodorou}(2006)}]{Tzoumanekas2006}
\bibinfo{author}{\bibfnamefont{C.}~\bibnamefont{Tzoumanekas}} \bibnamefont{and}
  \bibinfo{author}{\bibfnamefont{D.~N.} \bibnamefont{Theodorou}},
  \bibinfo{journal}{Macromolecules} \textbf{\bibinfo{volume}{39}},
  \bibinfo{pages}{4592} (\bibinfo{year}{2006}).

\bibitem[{\citenamefont{Hoy et~al.}(2009)\citenamefont{Hoy, Foteinopoulou, and
  Kr\"oger}}]{Hoy2009}
\bibinfo{author}{\bibfnamefont{R.~S.} \bibnamefont{Hoy}},
  \bibinfo{author}{\bibfnamefont{K.}~\bibnamefont{Foteinopoulou}},
  \bibnamefont{and} \bibinfo{author}{\bibfnamefont{M.}~\bibnamefont{Kr\"oger}},
  \bibinfo{journal}{Phys. Rev. E} \textbf{\bibinfo{volume}{80}},
  \bibinfo{pages}{031803} (\bibinfo{year}{2009}).

\bibitem[{\citenamefont{Everaers}(2012)}]{Everaers2012}
\bibinfo{author}{\bibfnamefont{R.}~\bibnamefont{Everaers}},
  \bibinfo{journal}{Phys. Rev. E} \textbf{\bibinfo{volume}{86}},
  \bibinfo{pages}{022801} (\bibinfo{year}{2012}).

\bibitem[{\citenamefont{Moreira et~al.}(2015)\citenamefont{Moreira, Zhang,
  M{\"u}ller, Stuehn, and Kremer}}]{Moreira2015}
\bibinfo{author}{\bibfnamefont{L.~A.} \bibnamefont{Moreira}},
  \bibinfo{author}{\bibfnamefont{G.}~\bibnamefont{Zhang}},
  \bibinfo{author}{\bibfnamefont{F.}~\bibnamefont{M{\"u}ller}},
  \bibinfo{author}{\bibfnamefont{T.}~\bibnamefont{Stuehn}}, \bibnamefont{and}
  \bibinfo{author}{\bibfnamefont{K.}~\bibnamefont{Kremer}},
  \bibinfo{journal}{Macromol. Theory Simul.} \textbf{\bibinfo{volume}{24}},
  \bibinfo{pages}{419} (\bibinfo{year}{2015}).

\bibitem[{\citenamefont{Doi}(1980)}]{Doi1980}
\bibinfo{author}{\bibfnamefont{M.}~\bibnamefont{Doi}}, \bibinfo{journal}{J.
  Polym. Sci. Polym. Phys. Ed.} \textbf{\bibinfo{volume}{18}},
  \bibinfo{pages}{1005} (\bibinfo{year}{1980}).

\bibitem[{\citenamefont{Lee and Kremer}(2009)}]{Lee2009}
\bibinfo{author}{\bibfnamefont{W.~B.} \bibnamefont{Lee}} \bibnamefont{and}
  \bibinfo{author}{\bibfnamefont{K.}~\bibnamefont{Kremer}},
  \bibinfo{journal}{Macromolecules} \textbf{\bibinfo{volume}{42}},
  \bibinfo{pages}{6270} (\bibinfo{year}{2009}).

\bibitem[{\citenamefont{Lee et~al.}(2010)\citenamefont{Lee, Halverson, and
  Kremer}}]{Lee2010}
\bibinfo{author}{\bibfnamefont{W.~B.} \bibnamefont{Lee}},
  \bibinfo{author}{\bibfnamefont{J.}~\bibnamefont{Halverson}},
  \bibnamefont{and} \bibinfo{author}{\bibfnamefont{K.}~\bibnamefont{Kremer}},
  \bibinfo{journal}{Macromolecules} \textbf{\bibinfo{volume}{43}},
  \bibinfo{pages}{3984} (\bibinfo{year}{2010}).

\bibitem[{\citenamefont{Sen et~al.}(2005)\citenamefont{Sen, Kumar, and
  Keblinski}}]{Sen2005}
\bibinfo{author}{\bibfnamefont{S.}~\bibnamefont{Sen}},
  \bibinfo{author}{\bibfnamefont{S.~K.} \bibnamefont{Kumar}}, \bibnamefont{and}
  \bibinfo{author}{\bibfnamefont{P.}~\bibnamefont{Keblinski}},
  \bibinfo{journal}{Macromolecules} \textbf{\bibinfo{volume}{38}},
  \bibinfo{pages}{650} (\bibinfo{year}{2005}).

\bibitem[{\citenamefont{Treloar}(1986)}]{Treloar1986}
\bibinfo{author}{\bibfnamefont{L.~R.~G.} \bibnamefont{Treloar}},
  \emph{\bibinfo{title}{The physics of Rubber Elasticity}}
  (\bibinfo{publisher}{Clarendon Press, Oxford}, \bibinfo{year}{1986}).

\bibitem[{\citenamefont{Halverson et~al.}(2013)\citenamefont{Halverson,
  Brandes, Lenz, Arnold, Bevc, Starchenko, Kremer, Stuehn, and
  Reith}}]{Espresso}
\bibinfo{author}{\bibfnamefont{J.~D.} \bibnamefont{Halverson}},
  \bibinfo{author}{\bibfnamefont{T.}~\bibnamefont{Brandes}},
  \bibinfo{author}{\bibfnamefont{O.}~\bibnamefont{Lenz}},
  \bibinfo{author}{\bibfnamefont{A.}~\bibnamefont{Arnold}},
  \bibinfo{author}{\bibfnamefont{S.}~\bibnamefont{Bevc}},
  \bibinfo{author}{\bibfnamefont{V.}~\bibnamefont{Starchenko}},
  \bibinfo{author}{\bibfnamefont{K.}~\bibnamefont{Kremer}},
  \bibinfo{author}{\bibfnamefont{T.}~\bibnamefont{Stuehn}}, \bibnamefont{and}
  \bibinfo{author}{\bibfnamefont{D.}~\bibnamefont{Reith}},
  \bibinfo{journal}{Comput. Phys. Comm.} \textbf{\bibinfo{volume}{184}},
  \bibinfo{pages}{1129} (\bibinfo{year}{2013}).

\end{thebibliography}

\end{document}